\documentclass{article}
\usepackage{graphicx}
\usepackage{bbm}
\usepackage{enumerate}
\usepackage{mathtools}
\usepackage{graphicx}%
\usepackage{multirow}%
\usepackage{amsmath,amssymb,amsfonts}%
\usepackage{amsthm}%
\usepackage{mathrsfs}%
\usepackage[title]{appendix}%
\usepackage{xcolor}%
\usepackage{textcomp}%
\usepackage{manyfoot}%
\usepackage{booktabs}%
\usepackage{algorithm}%
\usepackage{algorithmicx}%
\usepackage{algpseudocode}%
\usepackage{listings}
\usepackage[numbers]{natbib}
\usepackage{hyperref}

\newlength{\mylen}
\begin{document}
\title{FRODO: A novel approach to micro-macro multilevel regression}
\maketitle

\author{Shaun McDonald (Corresponding Author)}\footnotemark[1]

\author{Alexandre Leblanc}\footnotemark[2]

\author{Saman Muthukumarana}\footnotemark[2]

\author{David Campbell}\footnotemark[3]

\footnotetext[1]{Department of Statistics and Actuarial Science, Simon Fraser Univesity, 8888 University Drive, Burnaby, V5A 1S6, British Columbia, Canada. drshaunmcdonald at gmail dot com}
\footnotetext[2]{Department of Statistics, University of Manitoba, 66 Chancellors Circle, Winnipeg, R3T 2N2, Manitoba, Canada. alex.leblanc at umanitoba dot ca, saman.muthukumarana at umanitoba dot ca}
\footnotetext[3]{School of Mathematics and Statistics, Carleton University, 1125 Colonel By Drive, Ottawa, K1S 5B6, Ontario, Canada. davecampbell at math dot carleton dot ca}

\abstract{Within the field of hierarchical modelling, little attention is paid to micro-macro models: those in which group-level outcomes are dependent on covariates measured at the level of individuals within groups. Although such models are perhaps underrepresented in the literature, they have applications in economics, epidemiology, and the social sciences. Despite the strong mathematical similarities between micro-macro and measurement error models, few efforts have been made to apply the much better-developed methodology of the latter to the former. Here, we present a new empirical Bayesian technique for micro-macro data, called FRODO (Functional Regression On Densities of Observations). The method jointly infers group-specific densities for multilevel covariates and uses them as functional predictors in a functional linear regression, resulting in a model that is analogous to a generalized additive model (GAM). In doing so, it achieves a level of generality comparable to more sophisticated methods developed for errors-in-variables models, while further leveraging the larger group sizes characteristic of multilevel data to provide richer information about the within-group covariate distributions. After explaining the hierarchical structure of FRODO, its power and versatility are demonstrated on several simulated datasets, showcasing its ability to accommodate a wide variety of covariate distributions and regression models.}


\section{Introduction} \label{sec:frodo_intro}
Hierarchically structured data is quite common in statistics, with a litany of resources and methodology available for almost every imaginable configuration. Books such as \citep{Goldstein2010} provide comprehensive reviews on the subject of multilevel data. For the purposes of this manuscript, it will suffice to consider data organized in a two-level hierarchy. Data will be observed from ``groups'', each of which is comprised of multiple ``individuals'', with variables measured at either the group level (i.e.\ one measurement per group) or individual level (i.e.\ one measurement for each individual within each group).

Multilevel data structures can be broadly categorized into two types: \textit{macro-micro}, in which an individual-level outcome is predicted from group-level covariates; and \textit{micro-macro}, which is the opposite \citep{Snijders2011}. Although substantial attention has been given to the former structure (random effects models being one example of the macro-micro framework), the micro-macro paradigm is the subject of much less discussion \citep{FosterJohnson2018}, despite the occurrence of such datasets in health sciences \citep{Daouda2021}, sociology \citep{Bennink2013}, and economics \citep{Amendola2020}. Among the relatively few papers on the subject is the one by \citet{Croon2007}, one of the earliest papers to devise a method specifically for micro-macro regression. The data structure they considered (hereafter described as ``classical'') is as follows. Letting subscripts $i$ and $ij$ denote, respectively, the $i^\mathrm{th}$ group and the $j^\mathrm{th}$ individual within that group, the basic structure is
\begin{align}
Y_i &= \alpha + \beta \xi_i + \beta_Z Z_i + \epsilon_i, \label{eq:simple_reg} \\
X_{ij} &= \xi_i + \nu_{ij}. \label{eq:simple_covar}
\end{align}
Assuming group $i$ contains $n_i$ individuals, the observed data corresponding to that group is $\left\{Y_i, Z_i, X_{i1}, \ldots, X_{in_i}\right\}$. In words, $Y_i$ is a group-level response variable (with regression error $\epsilon_i$), $Z_i$ is a group-level scalar covariate, and the $X_{ij}$'s are individual-level measurements of some ``latent'' unobserved covariate $\xi_i$ with errors $\nu_{ij}$.
One can think of the model as two ``parts'': a regression part specified by (\ref{eq:simple_reg}), and a covariate observation part specified by (\ref{eq:simple_covar}). The linearity of the regression and additivity of the covariate error justify the ``classical'' moniker for this structure.

Although micro-macro modelling literature is relatively scarce, the structure implied by (\ref{eq:simple_reg}--\ref{eq:simple_covar}) is essentially equivalent to (a version of) the much better-studied \textit{classical measurement error model} \citep[chapter 1 of][and references therein]{Carroll2006}. The main difference is conceptual: in a micro-macro model, replicate covariate measurements correspond to distinct individuals within a group; while in a measurement error model, they are merely repeated noise-corrupted observations of some true explanatory variable for the $i^\mathrm{th}$ observational unit. There is another practical difference: most measurement error literature assumes smaller $n_i$'s (the number of covariate measurements per group) than one tends to encounter in a ``true'' micro-macro setting.

The simplest approach to modelling such data is the ``naive'' one: simply using the sample means $\bar{X}_{i} = n_i^{-1} \sum_j X_{ij}$ as proxies for the latent $\xi_i$'s. However, it is well-known \citep[e.g.\ chapter 3 of][and references therein]{Carroll2006} that such a failure to account for the uncertainty in the $X_{ij}$'s biases estimates of the regression parameters. Most notably, it creates \textit{attenuation} in the estimate of $\beta$: letting $\hat{\beta}$ denote such an estimate, we will have $\lvert \widehat{\beta} \rvert < \lvert \beta \rvert$, even as the number of groups grows asymptotically. In intuitive terms, this attenuation happens because the noise in the covariates stretches the regression line on the horizontal axis. Thus, a plethora of both frequentist and Bayesian methods have been proposed to account for covariate uncertainty in a way that produces less biased estimation and inference for the regression part of the model. A comprehensive review of measurement error methodology is beyond the scope of this manuscript, but the interested reader may refer to books such as \citep{Buonaccorsi2010,Carroll2006} or the review paper of \citet{Schennach2016}.

Many real-world datasets do not obey the ``classical'' framework of (\ref{eq:simple_reg}--\ref{eq:simple_covar}) \citep[e.g.\ Section 6.4 of][and references therein]{Buonaccorsi2010}, and there are two ways to transcend it: by replacing the linear terms $\beta\xi_i$ and $\beta_Z Z_i$ in (\ref{eq:simple_reg}) with arbitrary regression functions, or by generalizing the additive covariate structure in (\ref{eq:simple_covar}). There are few micro-macro modelling papers with generalizations of either type, aside from the discrete variable methods of \citet{Bennink2013, Bennink2016}. Thus, we focus our attention here on the measurement error literature instead. Beyond the comprehensive review sources mentioned above, the most generalized framework which is relevant to this manuscript is that of \citet{Hu2008}. They assumed each observational unit $i$ only has a single covariate measurement $X_i \sim f_{X \mid \xi = \xi_i}$, but also has a single replicate measurement or \textit{instrumental variable} $W_i$, assumed to provide further information about $\xi_i$. They also allowed a very general form for the regression function in which $Y$ only depends on the unobserved $\xi$, with only some technical assumptions on the distributions of $Y \mid \xi$, $X \mid \xi$, and $\xi \mid W$. Their assumptions on the covariate structure were very broad, requiring only that there exists a functional $M$ such that $M\left[f_{X \mid \xi}\left(\cdot \mid \xi\right)\right] \equiv \xi$ for all $\xi$. Examples of such functionals include the mode, as well as any quantile or moment. With this framework, the authors proposed a sieve likelihood estimator for the regression parameters and the densities of $X \mid \xi$ and $\xi \mid W$. To our knowledge, there are no established Bayesian methods that accommodate this level of generality. \citet{Sarkar2014} proposed a Bayesian model which used Dirichlet Process mixtures to achieve a great deal of flexibility in modelling the regression function, latent covariates, and error terms; but it still assumed an additive error structure of the form (\ref{eq:simple_covar}).

Neither of the aforementioned papers (or, indeed, any measurement error literature we have seen) gives much consideration to the ``unit-specific'' covariate distributions $f_{X\mid \xi = \xi_i}$ --- specifically, to any differences between them across units. This is understandable, as most errors-in-variables problems have no more than a single-digit number of covariate measurements available per unit, making any such differences irrelevant. However, in an explicitly multilevel setting, there are typically many more individuals per group \citep[e.g.][]{Croon2007, Amendola2020}, and it may be of interest to explicitly consider the group-specific covariate densities in inference. We believe that the Bayesian paradigm (or, at the very least, the empirical Bayesian paradigm) is the most natural setting in which to achieve this.

With all of the above considerations in mind, our goals in this paper are threefold. First, we seek to develop a(n empirical) Bayesian model with generality comparable to that of \citet{Hu2008}. Second, we wish to apply this model in the micro-macro multilevel setting, providing an ability to accommodate ``non-classical'' data structures which we believe is sorely missing in that literature. Our final goal is to leverage the data sizes characteristic of micro-macro situations in order to focus our inference not only on the regression part of the model, but also the distributions of ``individual-level'' covariates within each group.

To achieve these goals, we propose \textbf{FRODO} (Functional Regression On Densities of Observations), a method which unifies density estimation and functional regression in a joint empirical Bayesian model. Although the core idea of FRODO is a fairly straightforward combination of well-established methods in principle, it allows for a remarkable degree of generality in data structures, and its design proves to be far from trivial.

Before describing FRODO, we first give an overview of necessary functional data analysis concepts in Section \ref{sec:fda}. We then give a general overview of the FRODO model and its assumed data structure in Section \ref{sec:frodo}, followed by a detailed description of its prior and likelihood components, as well as its practical implementation. In Sections \ref{sec:frodo_sim} and \ref{sec:croon}, we show several simulation studies which demonstrate the potential generality of FRODO in both the regression and covariate observation parts of a micro-macro model.

\section{A brief review of key functional data analysis concepts} \label{sec:fda}
Broadly speaking, \textit{functional data analysis} (FDA) is a field of statistics in which the fundamental units of interest are (almost everywhere) smooth functions. A detailed overview of the field is beyond the scope of this manuscript, but the interested reader may find one in the excellent book by \citet{Ramsay2005}. Here we discuss only the concepts necessary to establish notation and motivation for FRODO.

\subsection{Scalar-on-function functional regression}
As the name implies, scalar-on-function regression concerns the modelling of a real-valued univariate (or ``scalar'') response variable with predictors that are functions \citep[][Section 12.3]{Ramsay2005}. This is achieved by using integrals in place of the sums which define scalar regression models. For example, consider a simple case in which our data are pairs $\left\{Y_i, f^*_i\right\}$, $i = 1, \ldots, N$, where $Y_i$ is a real-valued (continuous) scalar response and $f^*_i$ is an almost everywhere continuous function on $[0,1]$. For this data, a \textit{functional linear model} would be of the form
\begin{align}
Y_i = \alpha + \int_{0}^{1} \beta^*(x) f^*_i(x) \mathrm{d}x + \epsilon_i, \label{eq:func_reg}
\end{align}
with i.i.d.\ errors $\epsilon_i \sim \mathcal{N}\left(0, \sigma_Y\right)$. The \textit{coefficient function} $\beta^*$ weighting the integral is analogous to regression coefficients in a fully scalar regression model.

\subsection{Basis function expansions} \label{sec:basis}
Because function spaces are infinite-dimensional, a core component of FDA is the representation of functions of interest in finite-dimensional spaces \citep{Ramsay2005}. Typically, this is achieved by modelling functions as linear combinations of finitely many \textit{basis functions} \citep[][Section 3.3]{Ramsay2005}. Throughout this manuscript, we will use $f^*$ to denote a function of interest, and remove the asterisk to denote a relevant basis function approximation $f$.

Several types of functional bases exist, including those based on functional principal components, Fourier series, and splines \citep{Ramsay2005}. Attention here is restricted to the latter, and in particular the P-splines of \citet{Eilers1996}. For our purposes, it suffices to know that a P-spline representation of a function $f^*$ on a compact interval $[a, b]$ has the form
\begin{align}
f(x) =  \sum_{k=1}^K c_k B_{k}(x). \label{eq:direct_basis_frodo}
\end{align}
Here, the basis functions $B_{k}$ are splines: piecewise polynomials with supports defined by a set of equally-spaced ``knots'' in $[a, b]$. More detailed explanations of splines can be found in, for example, \citep{Ramsay2005} and \citet{Eilers1996}.
In the frequentist setting, the coefficients $c = \left(c_1, \ldots, c_K\right)$ can be fit with a penalized likelihood method. Common penalties force $f$ to adhere to desirable shapes by penalizing ``roughness'', as measured with a suitable linear differential operator \citep[see][Chapter 5]{Ramsay2005}. \citet{Eilers1996} modified this idea by instead using a penalty based on \textit{finite differences} between coefficients. Their penalty defines the notion of \textit{P-splines} and is of the form
\begin{align}
    \lambda \sum \limits_{k = r + 1}^{K} \left[\left(\Delta^r c \right)_{k - r}\right]^2, \label{eq:pspline_pen}
\end{align}
for a positive integer $r$, where $\Delta^r$ denotes the $r^\mathrm{th}$-order finite difference operator and $\left(\Delta^r c \right)_{k - r}$ denotes the $(k-r)^\mathrm{th}$ element of the $(K-r)$-dimensional vector $\left(\Delta^r c \right)$. For instance,
\begin{align}
\left(\Delta^1 c \right)_{1} &= c_2 - c_1, \nonumber \\ 
\left(\Delta^2 c \right)_{1} &= c_3 - 2c_2 + c_1, \textrm{\ and} \nonumber \\
\left(\Delta^3 c \right)_{1} &= c_4 - 3c_3 +  3c_2 - c_1. \nonumber
\end{align}
When the \textit{smoothing parameter} $\lambda > 0$ is large, (\ref{eq:pspline_pen}) dominates the penalized likelihood. Eilers and Marx noted that the sum in this penalty is a good approximation to the $r^\mathrm{th}$ derivative of $f$ when the knots defining the spline basis are equally spaced, especially for large dimensionality $K$. Thus, for large $\lambda$ the estimated $f$ is forced to take the approximate shape of a polynomial of degree $r-1$.

\citet{Lang2004} devised a Bayesian version of P-splines, based on the notion that a penalized likelihood function is analogous to a posterior distribution on the log scale, with the penalty term assuming the role of the prior. The penalty (\ref{eq:pspline_pen}) is the log density of an $r^\mathrm{th}$-order Gaussian \textit{random walk}:
\begin{align}
    \left(\Delta^r c\right)_{k-r} \sim \mathcal{N}\left(0, \frac{1}{\sqrt{2\lambda}}\right) \label{eq:RW}
\end{align}
for $k = r, r+1, \ldots, K$. \citet{Lang2004} gave the first $r$ components of $c$ (which we call ``\textit{free parameters}'' in contrast with the last $K-r$ components, whose behaviour is restricted by (\ref{eq:RW})) flat priors. However, we adopt the philosophy that such priors are unreasonable because they give equal weight to all values, no matter how extreme \citep[e.g.\ the case study of][]{Betancourt2017}, and we have also found such priors to result in extremely poor MCMC sampling behaviour in our models. Our priors on the free parameters in the various P-spline components of FRODO are described in Sections \ref{sec:dens_model}--\ref{sec:reg}.

As noted by \citet{Eilers1996}, one can use P-splines to model a density $f^*$ by replacing $f$ with $\log f$ in (\ref{eq:direct_basis_frodo}). The imposition of a polynomial shape on $\log f$ then leads to a density estimate which is close to the exponentiation of the corresponding polynomial. For instance, a penalty of order $r = 3$ (in either the frequentist or Bayesian setting) forces $\log f$ towards a quadratic shape, and therefore the resulting density estimate will be similar in shape to a Gaussian.

\section{The FRODO model} \label{sec:frodo}
\subsection{General overview} \label{sec:general}
Having reviewed the necessary functional data analysis concepts, we are now ready to describe the FRODO approach to micro-macro modelling. Assume the data is organized into $N$ groups, with the $i^\mathrm{th}$ group containing $n_i$ individuals. In the simplest case (assumed in the remainder of this section for ease of exposition), data in the $i^\mathrm{th}$ group consists of a group-level response variable $Y_i$, and individual-level observations of a covariate $X$, $\left(X_{i1}, \ldots, X_{in_i}\right)$. Although we assume real-valued Gaussian $Y_i$'s throughout this paper for the sake of simplicity, in principle the following methodology could be extended to any response type for which generalized linear modelling is possible. As in Section \ref{sec:frodo_intro}, the model is comprised of both a regression part and a covariate observation part, but we assume a much greater level of generality than in (\ref{eq:simple_reg}--\ref{eq:simple_covar}). Our only assumption for the covariate density part is that, for the $i^\mathrm{th}$ group, $X_i:= \left(X_{i1}, \ldots, X_{in_i}\right)$ (where an omitted subscript means the collection of all elements across that subscript) is an i.i.d.\ sample from an unobserved or ``latent'' group-specific covariate density $f^*_i$. The regression part of the model defines the ``novel'' idea at the core of FRODO: the use of these densities (technically, basis expansion estimators thereof) as predictors in a functional linear regression. In mathematical terms, the regression part of the model is
\begin{align}
Y_i &= \alpha + \int \beta^*(x) f_i^*(x) \mathrm{d} x + \epsilon_i \label{eq:frodo_int} \\
&= \alpha + \mathbb{E}^*_i\left[\beta^*\left(X\right)\right] + \epsilon_i, \label{eq:frod_gam} \\
\epsilon_i &\stackrel{i.i.d}{\sim} \mathcal{N}\left(0, \sigma_Y\right), \nonumber
\end{align}
where $\mathbb{E}^*_i\left[\beta^*\left(X\right)\right]$ denotes the expectation of $\beta^*\left(X\right)$ with respect to the density $f^*_i$. The equivalence between (\ref{eq:frodo_int}) and (\ref{eq:frod_gam}) is the key to FRODO's utility: by simply using densities as predictors in a functional linear regression, the resulting model is essentially a GAM. Thus, FRODO allows for a fully nonparametric approach to both regression functions and covariate structures.

It must be noted that the regressor in (\ref{eq:frod_gam}), $\mathbb{E}^*_i\left[\beta^*\left(X\right)\right]$, is the ``expectation of the regression function''. In general, this is \textit{not} equal to $\beta^*\left(\mathbb{E}^*_i\left[X\right]\right)$ --- the ``regression on the expectations'' --- unless $\beta^*$ is linear. Use of the latter is perhaps more ``standard'' in the measurement error literature, where it is typically assumed that the $X_{ij}$'s within each unit $i$ are noise-corrupted versions of some ``true'' covariate $\xi_i$ \citep[see][or any standard reference on measurement error]{Carroll2006}. Although it is not always assumed that $\mathbb{E}^*_i\left[X\right] = \xi_i$ (e.g.\ the general linear error structures described in Section 6.4 of \citep{Buonaccorsi2010}, and references therein), typically the target is estimation of $\beta^*\left(\xi_i\right)$, possibly marginalized over an estimate of the ``posterior'' $f_{\xi \mid X_i}$ \citep[e.g.][]{Hsiao1989, Li2002}. We are not aware of any literature which explicitly uses ``expectations of the regression'' in the way that FRODO does.

In the next two subsections, we detail the priors and likelihoods comprising FRODO. Recall that we approximate $\beta^*$ and the $f^*_i$'s with basis function expansions, use of which will be denoted without asterisks. In a slight abuse of notation, we consider the model
\begin{align}
Y_i &= \alpha + \int \beta(x) f_i(x) \mathrm{d} x + \epsilon_i \label{eq:frodo_int_basis} \\
&= \alpha + \mathbb{E}_i\left[\beta\left(X\right)\right] + \epsilon_i, \label{eq:frod_gam_basis}
\end{align}
as a proxy to (\ref{eq:frodo_int}--\ref{eq:frod_gam}), where $\beta$ and $f_i$ are the basis function approximations to their ``true'' counterparts ($\beta^*$ and $f^*_i$, respectively), and $\mathbb{E}_i$ denotes expectation w.r.t.\ $f_i$.

To formally justify this ``proxy model'' in terms of the ``true'' model (\ref{eq:frodo_int}), it would be necessary to replace the error terms in (\ref{eq:frodo_int_basis}) with terms that account for both the random regression error and the \textit{approximation error} inherent to such finite-dimensional approximations \citep[e.g.][]{Hansen2014}. Such terms would be of the form
\begin{align}
e_i = \epsilon_i + \int\left[\beta^*(x)f_i^*(x) - \beta(x)f(x)\right]\mathrm{d}x. \label{eq:combined_error}
\end{align}
However, in practice we effectively assume that the second term in (\ref{eq:combined_error}) is negligible and do not concern ourselves with the distinction between $\epsilon_i$ and $e_i$.

Before exploring the details of FRODO, some final technical and notational points are in order. We recommend standardizing the data so that default prior choices are weakly informative \citep[][Sections 2.9 and 16.3]{Gelman2013}. Keeping with our convention of using omitted subscripts to mean the collection of all elements across that subscript, let $Y = \left(Y_1, \ldots, Y_N\right)$ and $X = \left\{X_1, \ldots, X_N\right\}$, where $X_i$ was defined above. In what follows, we will assume that $Y$ and $X$ have both been standardized to have zero mean and unit variance. Note that for $X$, this standardization is ``marginal'', meaning that it is done across groups \textit{and} individuals within groups. We will overload notation and use $f^*_i$ and $f_i$ to refer to, respectively, the true density and its basis function approximation for the standardized version of $X_i$. For technical reasons, it is necessary to assume that $\beta$ and the $f_i$'s are all defined on a common compact interval. This will be denoted by $[a, b]$ on the standardized scale, and when it is necessary to speak about the domain of the covariates on the original (unstandardized) scale, it will be denoted by $[a', b']$. Assuming $X$ has been standardized as recommended above, we have $a = \left(a' - \bar{X}\right)/\sigma(X)$ and $b = \left(b' - \bar{X}\right)/\sigma(X)$, and $[a', b']$ can be chosen so that its endpoints are (nearly) equal to the unscaled extrema of the covariates.

\subsection{The density model} \label{sec:dens_model}
For computational convenience --- and because it suffices for the ordinal covariates which are common in real micro-macro datasets \citep[e.g.][]{Croon2007, Amendola2020, Daouda2021} ---  the $f_i$'s are modelled as histograms. In practical terms, this means that they are linear combinations of constant basis functions:
\begin{align}
f_i(x) &= \sum_{k=1}^K \phi_{ik} \mathbbm{1}_{I_k}(x), \label{eq:hist_dens}
\end{align}
where $I_k$ is the $k^\mathrm{th}$ equal-width subinterval $\left[a + (k-1)h, a + kh\right)$ of $[a, b]$, $\mathbbm{1}_{I_k}$ is the indicator function of $I_k$, and $h = (b-a)/K$ is the bin width. The density coefficients $\phi_{ik}$ are scaled ``softmax'' transformations of Gaussian random variables $\theta_{ik}, \, i = 1, \ldots, N, \, k = 1, \ldots, K$:
\begin{align}
\phi_{ik} = \frac{e^{\theta_{ik}}}{h\sum_{j=1}^K e^{\theta_{ij}}}, \label{eq:phi}
\end{align}
where, for all $i$, $\theta_{i1} \equiv 0$ to ensure identifiability. Equivalently, we may say that the $\phi_{i}$'s are (up to the scaling factor $h$) logistic normal random vectors \citep{Aitchison1980}.

The priors for the $\theta$'s are chosen in order to impose useful constraints on the behaviour of the densities. In particular, for some positive integer $r$ we will impose an $r^\mathrm{th}$-order Gaussian random walk prior on $\theta_i = \left(\theta_{i1}, \ldots, \theta_{iK}\right)$ for all $i$. Since the logarithms of the $f_i$'s are also piecewise constant, this structure means that $\log f_i$ is a Bayesian P-spline of degree zero, with $r^\mathrm{th}$-order penalty, for all $i$. Recall from Section \ref{sec:basis} that an $r^\mathrm{th}$-order random walk prior on $\theta_i$,
\begin{align}
\left(\Delta^r\theta_i\right)_{k-r} &\sim \mathcal{N}\left(0, \tau_i\right), \, k \geq r + 1 \label{eq:rand_walk}
\end{align}
forces $\log f_i$ towards the approximate shape of a $\left(r-1\right)^\mathrm{th}$-degree polynomial when the smoothing parameter $\tau_i$ is small\footnotemark.
\footnotetext{Henceforth, the phrase ``smoothing parameter'' will refer to the standard deviation of the random walk prior ($\tau$), instead of its precision as in Section \ref{sec:basis} (where it was denoted by $\lambda = \tau^{-2}/2$).}

Note that (\ref{eq:rand_walk}) completely determines the conditional distributions of $\theta_{ik}$ for $k > r$ given $\theta_{ik}$ for $k \leq r$. In the case $r > 1$, it remains to set the priors on the ``free parameters'' $\theta_{ik}$ for $2 \leq k \leq r$: the ``initial values'' of the random walk. A seemingly sensible and simple choice would be diffuse, mean-zero, independent Gaussian priors. Unfortunately, this turns out not to be entirely suitable for FRODO. For $r > 1$, imposing fully independent priors on the densities\footnotemark{} causes bias in the posterior mean coefficient function, $\widehat{\beta}$. For instance, if the true $\beta$ is a linear function, the magnitude of the slope of $\widehat{\beta}$ will be biased downward, just as in the ``naive'' approach to modelling described in Section \ref{sec:frodo_intro}. In the Bayesian hierarchical setting, this ``attenuation'' problem can be solved by putting priors on the covariates which introduce dependence between them and ``pool'' each group's measurements towards a latent group-level variable. The solution here is similar.
\footnotetext{When discussing the model itself, we will typically write ``the densities'' to refer to the histograms $f_i$ which are actually part of the model. When it is necessary to invoke the $f^*_i$'s, we will specify them as the ``\textit{true} densities''.}

To expand on this, first note that with $\theta_{i1} \equiv 0$ for all $i$, we have
\begin{align}
\theta_{ik} &= \log\left(h^{-1}\int\limits_{a + (k-1)h}^{a + kh} f_i(x) \mathrm{d} x\right) - \log\left(h^{-1}\int\limits_{a}^{a + h} f_i(x) \mathrm{d} x\right) \label{eq:true_diff} \\
&\approx \log f^*_i\left(a+ h\left(k -\frac{1}{2}\right)\right) - \log f^*_i\left(a + \frac{h}{2}\right), \label{eq:approx_diff}
\end{align}
recalling that $f^*_i$ is the ``true'' density for group $i$.

Suppose $f^*_i$ is that of a  $\mathcal{N}\left(\xi_i, \sigma_i\right)$ random variable\footnotemark. This corresponds to the limiting case for $r = 3$ as $\tau_i \to 0$, and it can be shown that (\ref{eq:approx_diff}) in this case reduces to
\footnotetext{Assuming the covariates have been standardized as recommended in Section \ref{sec:general}, most of $f^*_i$'s mass presumably lies in $[a, b]$, and $a < \xi_i < b$.}
\begin{align}
\theta_{ik} \approx \frac{h(k-1)}{\sigma_{i}^2}\left(\xi_i  - \left(a + \frac{kh}{2}\right)\right) \label{eq:free_approx_3}
\end{align}

For $r = 3$, this approximation motivates our choice of priors for the ``free parameters''. For each $i$ and $k = 2, 3$, we take them to be Gaussian with mean given by the right side of (\ref{eq:free_approx_3}) and standard deviation $\tau_i$. Thus, $\tau_i$ controls $f_i$'s adherence to the limiting Gaussian shape in two respects: by controlling the free parameters' deviations from their means, and by scaling the random walk behaviour in (\ref{eq:rand_walk}).

We now set priors on $\xi_i$ and $\sigma_i$. When the true covariate densities are Gaussian, the structure of the data is analogous to that of the ``classical'' micro-macro model, with $\xi_i$ being a ``latent group-level covariate'' and $\sigma_i$ controlling the level of Gaussian noise for each group's individual-level covariate measurements. In keeping with natural choices for that setting, we first assign the $\xi_i$'s a $\mathcal{N}\left(\mu_\xi, \sigma_\xi\right)$ prior. Recalling that $[a', b']$ denotes the assumed domain of the covariate densities on the original (unstandardized) scale, the mean $\mu_\xi$ is given a $\mathcal{N}\left((a'b - b'a)/(a'-b') , 15/K^2\right)$ hyperprior. This corresponds to a mean-zero hyperprior on the original covariate scale, with the empirically-determined standard deviation $15/K^2$ accounting for the discretization error from approximation (\ref{eq:approx_diff}). The scale $\sigma_\xi$ is given a standard half-normal prior, which will be fairly uninformative if the $X_{ij}$'s have been scaled to have unit marginal variance. It will often be reasonable that the covariate densities are homoscedastic: $\sigma_{i} \equiv \sigma_X$ for all $i$. A standard half-normal prior is a sensible choice in this case. If one wishes to explicitly model heterogeneity, then each $\sigma_{i}$ can be given its own half-normal prior, perhaps sharing a common scale parameter with its own hyperprior.

Now, suppose $f^*_i$ is instead a (shifted) $\mathrm{Exponential}\left(\lambda_i\right)$ density. This corresponds to the limiting case for the random walk with $r = 2$, and here (\ref{eq:approx_diff}) reduces to
\begin{align}
\theta_{ik} = -\lambda_i\left(k-1\right)h. \label{eq:exp_free}
\end{align}
Note that for an exponential density, there is no discretization error, so (\ref{eq:true_diff}) and (\ref{eq:approx_diff}) are equal. Thus, analogously to the $r = 3$ case described above, when $r = 2$ we assume the ``free parameters'' $\theta_{i2}$ are Gaussian with mean given by the right side of (\ref{eq:exp_free}) and standard deviation $\tau_i$.  A natural choice of prior for the ``latent rates'' $\lambda_i$ is $\mathrm{Gamma}\left(\alpha_\lambda, \alpha_\lambda/\mu_\lambda\right)$. The mean $\mu_\lambda$ is given a standard half-normal prior (which should be only weakly informative if the covariates have been standardized), while the shape parameter $\alpha_\lambda$ is given a more diffuse half-normal prior with scale 10. Note that this parameterization of the Gamma in terms of shape and mean, rather than the more conventional shape and rate, proved computationally advantageous.

By defining the ``free parameters'' in terms of latent group-level variables with their own hyperpriors, we introduce the necessary dependence and ``pooling'' to prevent bias in the regression part of the model, just as one might do in the scalar case.  For any order $r$, the density model is completed with priors on the smoothing parameters $\tau_i$, which we take to be exponentials with rates $\delta_i^{-1}$. The scales are assumed to be fixed data, chosen empirically based on heuristics and the properties of the $X_{ij}$'s in the absence of more meaningful prior information. Such choices place FRODO in the category of ``empirical Bayesian'' methods, but we have found that sampling behaviour and posterior results can become poor when the $\delta_i$'s are not chosen carefully.
If group sizes are moderate ($n_i$'s roughly between 20 and 60) and one doesn't expect any of the covariate densities to deviate too seriously from the shape implied by the $r^\mathrm{th}$-order random walk prior, $\delta_i = 0.1$ for all $i$ seems to be a good default choice based on preliminary empirical results. Smaller groups tend to require smaller $\delta_i$'s, and it may also be advantageous to shrink them when the basis dimension $K$ is very large, especially relative to the $n_i$'s.

Finally note that, because the densities are piecewise constant, the likelihood $X_i \sim f_i$ is equivalent to $m_i := \left(m_{i1}, \ldots, m_{iK}\right) \sim \textrm{Multinomial}\left(n_i, \phi_i\right)$, where $m_{ik}$ is the bin count $\left|\left\{j: X_{ij} \in I_k\right\}\right|$. In summary, the model for the densities, assuming an $r^\mathrm{th}$-order random walk prior structure (for $r \leq 3$), is
\begin{equation}
  \settowidth{\mylen}{$\displaystyle \theta_{i2} \sim {}$\,}
  \begin{aligned}
m_i &\sim \textrm{Multinomial}\left(n_i, \phi_i\right) \nonumber \\
\phi_{ik} &= \frac{e^{\theta_{ik}}}{h\sum_{j=1}^K e^{\theta_{ij}}} \nonumber \\
\theta_{i1} &\equiv 0 \nonumber \\
\\
      &\phantom{{}={}}
      \hspace*{-\dimexpr\mylen+\nulldelimiterspace}
        \left.\begin{aligned}
          \theta_{i2} &\sim \mathcal{N}\left(-\lambda_i h, \tau_i\right) \\
             \lambda_i &\sim \mathrm{Gamma}\left(\alpha_\lambda, \frac{\alpha_\lambda}{\mu_\lambda}\right) \nonumber \\
\alpha_\lambda &\sim \text{Half-Normal}(0,10) \nonumber \\
\mu_\lambda &\sim \text{Half-Normal}(0,1)  \nonumber
        \end{aligned}\right\}
        \quad r = 2 \\
\\
      &\phantom{{}={}}
      \hspace*{-\dimexpr\mylen+\nulldelimiterspace}
        \left.\begin{aligned}
          \theta_{ik} &\sim \mathcal{N}\left(\frac{h(k-1)}{\sigma_{i}^2}\left(\xi_i  - \left(a + \frac{kh}{2}\right)\right), \tau_i\right)\ (k = 2, 3) \\
             \xi_i &\sim \mathcal{N}\left(\mu_\xi, \sigma_\xi\right) \nonumber \\
\mu_\xi &\sim \mathcal{N}\left(\frac{a'b - b'a}{a'-b'}, \frac{15}{K^2}\right) \\
\sigma_\xi &\sim \text{Half-Normal}(0, 1) \nonumber \\
\sigma_X &\sim \text{Half-Normal}(0,1)  \nonumber
        \end{aligned}\right\}
        \quad r = 3 \\ \\
\left(\Delta^r\theta_i\right)_{k-r} &\sim \mathcal{N}\left(0, \tau_i\right), \, k > r \nonumber \\
\tau_i &\sim \text{Exp}(\delta_i^{-1}) \nonumber \\
  \end{aligned}
\end{equation}

\subsection{The regression model} \label{sec:reg}
Here we detail priors for the regression part of FRODO, the likelihood for which is defined by (\ref{eq:frodo_int_basis}--\ref{eq:frod_gam_basis}). Recall that we have restricted our attention in this manuscript to continuous real-valued responses $Y_i$ with i.i.d.\ errors $\epsilon_i \sim \mathcal{N}\left(0, \sigma_Y\right)$, The following priors on $\alpha$ and $\beta$ would require only minor changes to accommodate more general response types (e.g.\ different scaling may be in order to ensure plausible effect sizes in a logistic regression; see Section 16.3 of \citet{Gelman2013}), and the prior on the dispersion parameter could easily be changed as necessary.

The error scale $\sigma_Y$ is given a half-T prior with 4 degrees of freedom and scale $1/\sqrt{2}$, so that $\sigma_Y$ has a prior mean of $1/\sqrt{2}$. Recalling the assumption from Section \ref{sec:general} that $Y$ has been standardized to have unit variance, this scale (in informal terms) loosely corresponds to a prior expectation that roughly half of the variation in the response values is due to regression error (assuming that the errors and regressors are independent, which we do here). This seems to be a sensible approach for a ``default'' prior, unless one has prior domain knowledge which would allow for context-specific prior beliefs about the regression error.

Both $\alpha$ and $\beta$ are given hierarchical priors with scales proportional to $\sigma_Y$. This can be shown to ensure unimodality in some penalized Bayesian regression models \citep{Park2008}, and we also found that it improved sampling behaviour. The intercept $\alpha$ is given a diffuse $\mathcal{N}\left(0, 20\sigma_Y\right)$ prior.

We take the coefficient function $\beta$ to be piecewise constant, with the same dimensionality $K$ as the densities. This is quite computationally convenient, as the integral in (\ref{eq:frodo_int_basis}) then reduces to the inner product between the coefficients of $\beta$ and $f_i$, scaled by the bin width $h$. Because the functional predictors all have unit integral, adding a constant shift to $\beta$ does not change the model: for any $c \in \mathbb{R}$, the model is identical if $\beta$ and $\alpha$ are replaced by $\beta + c$ and $\alpha - c$, respectively. Thus, we impose the identifiability constraint $\mathbb{E}\left[\beta\left(X\right)\right] := \int_a^b \widehat{f}_\mathrm{Cent}(x) \beta(x) \mathrm{d}x = 0$, where $\widehat{f}_\mathrm{Cent}$ is the \textit{empirical central density}:
\begin{align}
\widehat{f}_\mathrm{Cent}(x) := \sum_{k = 1}^K \frac{\sum_{i = 1}^N m_{ik}}{\sum_{l = 1}^K \sum_{i = 1}^N m_{il}}\mathbbm{1}_{I_k}(x).
\end{align}
Essentially, $\widehat{f}_\mathrm{Cent}$ is the ``marginal histogram'' of all covariate data across groups. Presumably, the total number of covariate observations $\sum_i n_i$ will be large enough in most data sets to ensure that $\widehat{f}_\mathrm{Cent}$ is reasonably ``smooth'', so that it is a good approximation to the ``marginal'' covariate density (i.e.\ marginalized across groups) for large $K$. Note that we use the \textit{empirical} central density mainly for computational convenience: an ``inferred central density'' like $N^{-1}\sum_i f_i$ would certainly be ``smoother'', but this would add needless complexity to the gradients used in NUTS when the empirical version is sufficient to ensure identifiability.

This constraint amounts to centering the inferred regressors $\mathbb{E}_i\left[\beta(X)\right]$. In practice, the constraint is achieved by defining a piecewise constant function
\begin{align}
\beta^0(x) & := \sum_{k=1}^K \beta^0_{k} \mathbbm{1}_{I_k}(x) \label{eq:beta0}
\end{align}
and taking $\beta = \beta^0 - \int \widehat{f}_\mathrm{Cent} \beta^0$. In keeping with Bayesian functional regression approaches such as \citep{Crainiceanu2010}, we put a second-order random walk prior on the coefficients of $\beta^0$, with the first coefficient set to 0 for identifiability:
\begin{align}
\beta^0_{1} &\equiv 0 \nonumber, \\
\beta^0_2 &\sim \mathcal{N}\left(0, 20h\sigma_Y\right) \nonumber, \\
\left(\Delta^2\beta^0\right)_{k-2} &\sim \mathcal{N}\left(0, \tau_\beta \sigma_Y \right). \nonumber
\end{align}
The smoothing parameter $\tau_\beta$ controls the extent to which $\beta$ deviates from the random-walk behaviour. As $\tau_\beta \to 0$, $\beta$ is forced towards a stepwise approximation to a straight line, and the regression model (\ref{eq:frodo_int_basis}) is therefore forced towards a linear regression. In this limiting case, the ``slope'' of $\beta$, $h^{-1}\beta^0_2$, is equivalent to the regression coefficient in a scalar linear model. Thus, using a scale factor of $20\sigma_Yh$ in $\beta^0_2$'s prior can be considered roughly analogous to placing a $\mathcal{N}\left(0, 20\sigma_Y\right)$ prior on the coefficient in the scalar case, which should be reasonably diffuse if the covariates have been scaled as recommended above \citep[e.g][Section 25.12 of User's Guide]{stanref}. Finally, $\tau_\beta$ is given an exponential prior with rate 2 (equivalently, scale 0.5). In contrast to the smoothing parameters for the densities, we found that $\tau_\beta$ did not require a careful selection of prior scale in order to ensure good model performance. 

\subsection{Implementation} \label{sec:imp}
The FRODO model is implemented in the Stan programming language \citep{Carpenter2017}, which provides exceptional power, flexibility, and efficiency through its use of the No-U-Turns Sampling (NUTS) variant of Hamiltonian Monte Carlo \citep{Hoffman2014}. For each of the below simulation studies, four parallel chains were run with fairly diffuse starting values, with sufficiently many sampling iterations to ensure effective sample sizes of at least 450 for all parameters \citep[see][Section 11.5]{Gelman2013}. All model runs were devoid of divergent transitions \citep{stanref}, and the overwhelming majority of parameters in all simulations had $\hat{R}$ values (where $\hat{R}$ is a diagnostic which helps to assess model convergence, see \citet{Vehtari2021}) below 1.01, with only a single parameter in each of the models of Sections \ref{sec:normal_quad} and \ref{sec:beta_quad} having a value very slightly above this threshold. All of the simulation studies below were conducted using R \citep{R}, interfacing with Stan via the RStan package \citep{rstan}. More details are given in Appendix \ref{chapter:app_frodo}.

\section{Simulation studies} \label{sec:frodo_sim}
As discussed in previous sections, FRODO is uniquely powerful in theory because it is ``doubly nonparametric'': it can capture arbitrary unknown structures in both the covariate densities and the regression model. In the following subsections, we put this to the test with a wide variety of simulated datasets. We will assess FRODO's ability to harness location, scale, and shape information from covariate densities and use it to recover true regression relationships. In each study, FRODO will be compared to two simpler models:
\begin{enumerate}
    \item a ``naive'' scalar regression model using only the sample means of the covariate measurements (or of some suitable transformation thereof, where applicable); and
    \item a ``hierarchical'' scalar regression model, where the form of the regression function and covariate distributions are assumed known, with only the actual parameter values unknown.
\end{enumerate}
More detail will be provided in the following subsections.

Because FRODO does not assume any parametric form for either the regression or covariate parts of the model, all that is required are choices of an appropriate random walk order $r$, dimensionality $K$, (unstandardized) density domain $[a',b']$, and set of density scaling factors $\delta = \left(\delta_1, \ldots, \delta_N\right)$. These choices must be made assuming that the true data-generating mechanisms are not known \textit{a priori}. One could use subject-specific domain knowledge if it is available. Otherwise, an ``empirical Bayesian'' approach based on informal inspections of the data is acceptable, and this is the approach we will use for all simulation studies in this manuscript. Visual inspection of default histograms or KDE's suffices to this end. From a strictly Bayesian perspective on inference, one could argue that this data dependence in the prior is not philosophically sound. However, an empirical Bayesian approach to nonparametric modelling is certainly not without precedent \citep[e.g.][]{Rousseau2016, VandeWiel2019}. \citet{Serra2017} devised an empirical Bayesian method for determining both the smoothing parameter and penalty order in spline fitting; our strategy could be viewed as a crude, heuristic approximation of such a method.

\subsection{Gaussian covariate densities, linear regression model} \label{sec:gauss_lin}
We begin with the ``classical'' structure from Section \ref{sec:frodo_intro}, where the individual-level measurements within groups are Gaussian deviations from a latent group-level covariate, itself Gaussian:
\begin{align}
\xi_i & \sim \mathcal{N}\left(0, \sigma_\xi\right), \label{eq:covar} \\
X_{ij} &\sim \mathcal{N}\left(\xi_i, \sigma_{X}\right). \label{eq:obs}
\end{align}
The regression model is also linear:
\begin{align}
Y_i &= \alpha + \tilde{\beta} \xi_i + \epsilon_i, \label{eq:reg} \\
 &= \alpha + \mathbb{E}_i\left[\tilde{\beta} X\right] + \epsilon_i, \nonumber \\
\epsilon_i &\sim \mathcal{N}\left(0, \sigma_Y\right). \label{eq:err}
\end{align}
Note that the second line explicitly restates the regression model in the form of (\ref{eq:frod_gam}), with the true regression function $\beta^*(x) = \tilde{\beta} x$ being a line with slope $\tilde{\beta}$. Some clarification on notation is in order here. Throughout Sections \ref{sec:frodo_sim}--\ref{sec:croon}, $\tilde{\beta} \in \mathbb{R}$ will denote a scalar which determines the magnitude and sign of the true regression function $\beta^*$. In turn, recall that the (piecewise constant) basis function approximation to $\beta^*$ is denoted as $\beta$.

The true parameter values \textit{before} standardizing\footnotemark{}\footnotetext{Throughout this section, all parameter values and results will be presented on the original (unstandardized) scale of the given data. The standardization only occurs ``internally'', during the fitting of the FRODO model.} the data as described in Section \ref{sec:dens_model} are $\sigma_\xi = 2$, $\sigma_X = 3$, $\alpha = 0.3$, $\tilde{\beta} = 0.4$, and $\sigma_Y = 0.5$. The result is a dataset with moderate amounts of noise in both the regression and the covariate measurements. The number of groups is $N = 275$ and each group contains covariate samples for $n = 20$ individuals.

Upon inspecting the data as recommended in the introduction to this section (not shown), we find that an assumption of roughly Gaussian density shape (corresponding to $r = 3$) is reasonable for these data. Because the densities are moderately wide but relatively close together (as the between-density variability $\sigma_\xi$, is somewhat smaller than the within-density variability, $\sigma_X$), a modest basis of size $K = 10$ should suffice without substantial loss of information. For this simulated data we have $\min_{i,j} X_{ij} = -13.54922$, $\max_{i,j} X_{ij} = 10.87845$, so we extend this range slightly by the same amount in each direction to arrive at an assumed density domain\footnotemark{} of $[a', b'] = [-13.67077, 11]$. Finally, the default choice of $\delta_i = 0.1$ for all $i$ recommended in Section \ref{sec:dens_model} is used here.
\footnotetext{Henceforth, the ``assumed domain'' will be stated on the unstandardized scale of the original data (i.e.\ $[a', b']$), with the standardization to $[a,b]$ left unstated.}

As stated at the beginning of this section, we compare FRODO to two simpler models. The first is simply a standard Bayesian linear regression, with (\ref{eq:covar}) omitted and the group-level sample covariate means $\bar{X}_i$ treated as the ``true'' covariates.
The second is a scalar micro-macro Bayesian regression, implemented in the ``obvious'' way: namely, (\ref{eq:covar})--(\ref{eq:err}) are assumed to be the known form of the model, with all parameters (including the latent $\xi_i$'s) unknown and inferred.  Recall that the estimate of $\tilde{\beta}$ from the ``naive'' model will be smaller in magnitude than the ``true'' value, which the hierarchical scalar model will presumably recover more effectively.

\begin{figure}
\centering{\includegraphics[width=\textwidth]{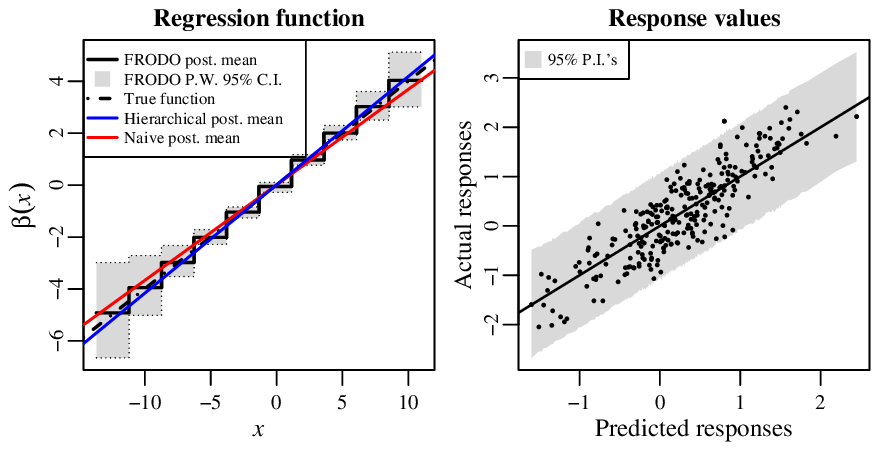}}
\caption{Results of FRODO applied to data with Gaussian covariates and a linear regression structure. Left: the regression function estimated by FRODO, alongside its pointwise 95\% credible region, the true function, and posterior mean estimates from hierarchical and naive scalar models. Right: responses $\hat{Y}_i$ predicted by FRODO (along with 95\% prediction intervals) vs.\ true responses.} \label{fig:norm_linear_reg}
\end{figure}

Figure \ref{fig:norm_linear_reg} shows results for the regression part of the model. In the left plot, the stepwise estimator of the regression function is shown with its pointwise (P.W.)\ 95\% credible interval (C.I.). Superimposed on the plot are the true regression function, as well as the posterior means from the hierarchical and naive scalar models (both of which assume a known linear form for the regression unlike FRODO, which only controls adherence to a linear regression through $\tau_\beta$). Because the within-group variability is not too much larger than the across-group variability and the sample sizes are reasonable, only a small amount of attenuation is caused by using the naive model, so the estimated regression functions for both scalar models are entirely within the pointwise C.I. from FRODO. However, the ``slope'' of the mean regression function from FRODO seems to be closer to those of the true function and the hierarchical scalar estimate, rather than that of the naive estimate. We can formalize this observation by considering the secant line to the FRODO regression function which intersects it at the midpoints of the first and last bins. The slope of this line (which is roughly analogous to a notion of ``slope'' for the FRODO regression function) is 0.4002, whereas the slopes of the true, hierarchical scalar, and naive scalar regressions are 0.4, 0.4172, and 0.3678, respectively.

Another way to assess FRODO's ability to infer the ``true'' regression (rather than the incorrect one implied by the naive model) is by checking the posterior for the regression error scale, $\sigma_Y$. Because of the additional noise in the individual-level covariate measurements, the naive model's estimate for $\sigma_Y$ will be biased upward \citep[e.g.][Section 3.2.1]{Carroll2006}. Indeed, the posterior mean for this parameter from the naive scalar model is 0.5556 (95\% C.I. $(0.5107, 0.6027)$), while the posterior means from FRODO and the hierarchical scalar model are 0.4930 (95\% C.I. $(0.4421, 0.5473)$) and 0.4904 (95\% C.I. $(0.4384, 0.5453)$), respectively. Because the FRODO estimate is much closer to the true value of 0.5 than it is to the ``naive estimate'', we are satisfied that we have avoided the attenuation problem inherent in the naive model. Table \ref{tab:sigma_y} contains summaries of the $\sigma_Y$ posteriors for every simulation study in this manuscript.

On the right of Figure \ref{fig:norm_linear_reg}, we have plotted the posterior mean predicted responses $\hat{Y}_i$ against the observed responses. The shaded region is a visual representation of 95\% posterior prediction intervals (P.I.'s)\ for each group.

\begin{figure}
\centering\includegraphics[width=\textwidth]{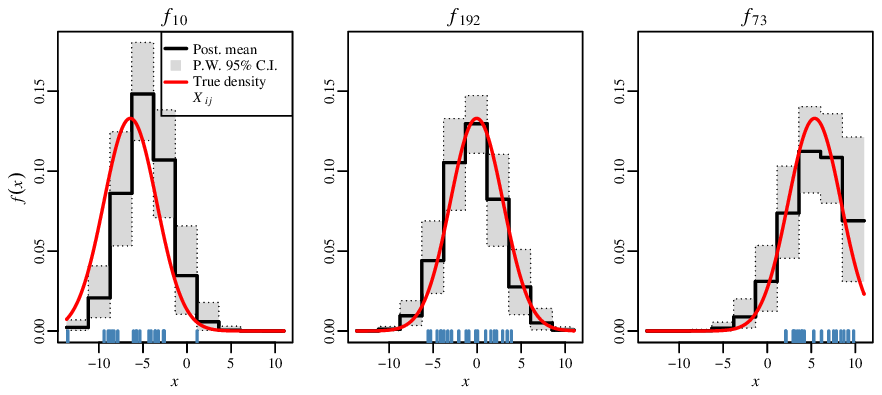}
\caption{For a selection of groups (from the data with Gaussian covariates and a linear regression structure), the FRODO estimate of the group-specific covariate density, alongside its pointwise 95\% credible regions. The true densities are superimposed as red lines, and the actual covariate samples are shown as rug plots.} \label{fig:normal_linear_densities}
\end{figure}

Figure \ref{fig:normal_linear_densities} shows the estimated $f_i$'s, along with their pointwise 95\% C.I.'s, for the group with the smallest (left) and largest (right) $\xi_i$'s, as well as the group whose $\xi_i$ is closest to the sample mean (middle). The middle and right fits are satisfactory, with the inference effectively capturing the true covariate densities (shown in red). The left plot shows that there is something of a mismatch between the inferred and true densities for the group with the lowest $\xi_i$, with the former shifted slightly too far to the right. Given that the model appears to perform well in all other respects, this is not a significant concern, especially since the rug plot suggests consistency with the data. We did not observe this problem in other datasets generated with the same parameter values (not shown), and therefore assume it is simply an unfortunate quirk of this particular data.

\subsection{Gaussian covariate densities, nonlinear regression model} \label{sec:normal_quad}
Here, we test FRODO's ability to handle nonlinear regression functions. The covariates adhere to the same Gaussian structure as in Section \ref{sec:gauss_lin}, but the regression model is now quadratic:
\begin{align}
Y_i &= \alpha + \tilde{\beta}\left(\xi_i^2 + \sigma_X^2\right) + \epsilon_i \nonumber \\
 &= \alpha + \mathbb{E}_i\left[\tilde{\beta} X^2\right] + \epsilon_i. \nonumber
\end{align}
Because the true covariate densities all have common variance $\sigma_X^2$, the difference between $\mathbb{E}_i\left[X^2\right]$ and $\left(\mathbb{E}_i\left[X\right]\right)^2$ is constant and can therefore be absorbed into the intercept. Here, the regression function is $\beta^*(x) = \tilde{\beta} x^2$.

The same parameter values $\left(\sigma_\xi, \sigma_X, \alpha, \tilde{\beta}, \sigma_Y\right) = \left(2, 3, 0.3, 0.4, 0.5\right)$ and number of groups $N = 275$ are used as in Section \ref{sec:gauss_lin} although the data is not strictly the same as we used a different seed for pseudorandom number generation in this study. Because the values of $\xi^2$ span a wider interval than those of $\xi$, the ``relative'' level of regression error is lower than in Section \ref{sec:gauss_lin}, since the ``signal'' is larger in scale than the ``noise''.

As before, we compare FRODO to two scalar models, one hierarchical and one naive. Here, however, it is assumed known in the hierarchical model that the regression is quadratic in the latent covariates $\xi$, with no linear term. The naive scalar model here is a GAM rather than a linear model, with the covariates taken to be the group-level sample means and the unknown regression function modelled as a cubic P-spline with second-order penalty.

Because the regression function is not one-to-one, an interesting difficulty arises in this framework when the group sizes $n_i$ are too small. On the regression side, the distributions are unchanged if $\xi_i$ is replaced with $-\xi_i$ in a given group. When $n_i$ is small and the true $\xi_i$ is close to zero, the available measurements $X_i$ may not be informative enough to distinguish between these possibilities\footnotemark. This creates multimodality in the posterior (for the hierarchical scalar model, and for FRODO to a somewhat lesser extent) with all of its associated difficulties, including poor HMC sampling behaviour and posterior mean estimates that are not particularly meaningful. Thus, larger group sizes are required if one wants meaningful inference on the covariate parameters as well as the regression parameters. Here, we increase the group size in the simulated data from the $n = 20$ used in Section \ref{sec:gauss_lin} to $n = 50$ for all $i$. The $X_{ij}$'s range from -13.76074 to 14.0043, and we expand this range by a small amount in each direction for an assumed density domain of $[-13.80644, 14.05]$. As before, we find $K = 10$ and $\delta_i = 0.1\ \forall i$ to be suitable choices here.
\footnotetext{This appears to also depend on the amount of covariate variability withing the group relative to the size of its regression error, although it is not currently clear exactly how this dependence works.}

\begin{figure}
\centering{\includegraphics[width=\textwidth]{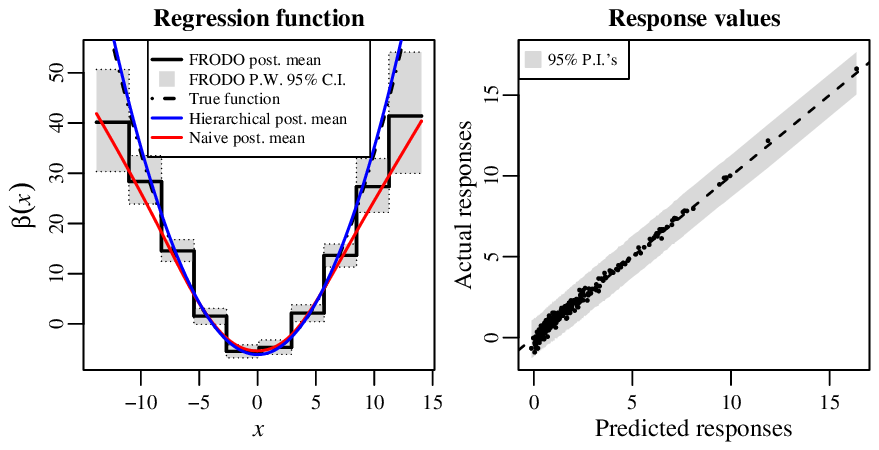}}
\caption{Results of FRODO applied to data with Gaussian covariates and a quadratic regression structure. Left: the regression function estimated by FRODO, alongside its pointwise 95\% credible region, the true function, and posterior mean estimates from hierarchical and naive scalar models. Right: responses $\hat{Y}_i$ predicted by FRODO (along with 95\% prediction intervals) vs.\ the true response values.} \label{fig:norm_square_reg}
\end{figure}
Results for the regression part of the model are shown in Figure \ref{fig:norm_square_reg}. At first glance, it may appear as though the FRODO estimate of the regression function is too attenuated, as it is closer to the estimate from the naive scalar model at the endpoints than it is to the true function and the hierarchical scalar estimate. Note, however, that over 95\% of the $X_{ij}$'s lie within the middle six bins, and over 95\% of the true latent $\xi_i$'s within the middle four. In those regions, the FRODO estimate is quite close to the true quadratic regression function. Towards the endpoints where the $X_{ij}$'s are very sparse, there is much less information with which to estimate value of the regression function. This edge effect is readily seen in several examples in this manuscript by observing that the pointwise C.I.'s for $\beta$ are wider in regions with few covariate estimates. In the linear example of Section \ref{sec:gauss_lin}, this did not create noticeable bias in the actual posterior mean for $\beta$ near the endpoints. Presumably this is because --- in somewhat informal terms --- the covariates in the middle of the domain were sufficiently informative to constrain the posterior for $\beta$ to a linear shape with high probability, which results in the smoothing parameter $\tau_\beta$ being small with high probability, which, in turn, enforces a linear shape in $\beta$ with fairly high probability throughout the rest of the domain. In this example, we do not penalize $\beta$ towards a quadratic shape --- only away from a linear shape. As such, it is not surprising that the posterior for $\beta$ is biased away from the truth near the endpoints, as neither the prior nor the likelihood are very informative there. In principle, one could specify a third-order random walk prior for $\beta$ in order to ensure a more genuinely quadratic shape, provided one had sufficient reason \textit{a priori} to assume this was an appropriate choice. However, we argue that the second-order random walk prior used here is more intuitive, as it is formulated in terms of deviations from a linear model. At any rate, the heightened bias and uncertainty in the FRODO regression function near the endpoints does not create any seriously adverse consequences for the rest of the inference. In particular, the FRODO posterior mean for $\sigma_Y$ is 0.4734 (95\% C.I. $(0.3849, 0.5643)$), much closer to the true value of 0.5 than the estimate from the naive model (0.8863, 95\% C.I. $(0.8152, 0.9669)$), suggesting that FRODO is successfully recovering the true regression model and not the biased naive version. The plot of estimate vs.\ true responses on the right of Figure \ref{fig:norm_square_reg} shows an overall good fit, although there is a small amount of bias in the estimates of the lowest responses.

\begin{figure}
\centering\includegraphics[width=\textwidth]{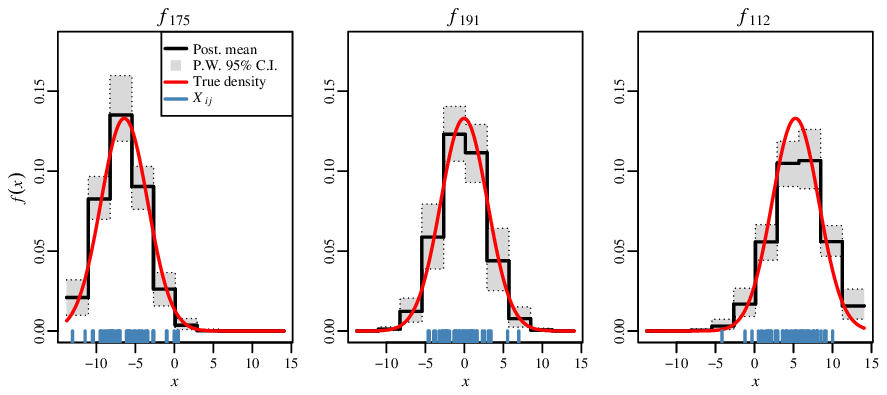}
\caption{For a selection of groups (from the data with Gaussian covariates and a quadratic regression structure), the FRODO estimate of the group-specific covariate density, alongside its pointwise 95\% credible region. The true densities are superimposed as red lines, and the actual covariate samples are shown as rug plots.} \label{fig:norm_square_dens}
\end{figure}
Figure \ref{fig:norm_square_dens} shows a sample of covariate densities, once again for the group with the smallest and largest $\xi_i$'s, and the $\xi_i$ closest to the sample mean. With larger group sizes, FRODO successfully approximates the true densities for each group shown here.

\subsection{Exponential covariate densities, linear regression model} \label{sec:exp_linear}
Although it is useful to model arbitrary regression functions, doing so with Gaussian covariate distributions is a capability shared by many methods. In fact, authors such as \citet{Sarkar2014} have developed Bayesian methods which allow for even more general structures of the form $X_{ij} = \xi_i + \nu_{ij}$. The true advantage of FRODO lies in its ability to handle covariates that are not based on any kind of \textit{additive} error structure. To demonstrate this, here we use an exponential covariate structure:
\begin{align}
\lambda_i & \sim \textrm{Gamma}\left(10, 10\right) \nonumber \\
X_{ij} &\sim \textrm{Exponential}\left(\lambda_i\right);\nonumber
\end{align}
and a linear regression model
\begin{align}
Y_i &= \alpha + \tilde{\beta}\lambda_i^{-1} + \epsilon_i \nonumber \\
&= \alpha + \mathbb{E}_i\left[\tilde{\beta} X\right] + \epsilon_i, \nonumber
\end{align}
where we have not restated the distribution for the error variance since it is identical to (\ref{eq:err}) for all subsequent studies.

It is worth contrasting this framework with that of Section \ref{sec:gauss_lin}. There, the true covariate densities were Gaussians with equal variances, so the group-level responses depended on their \textit{locations}. With exponential covariate distributions, the linear regression model implies responses that instead vary with the \textit{scales} of the densities. This turns out to be a somewhat challenging type of model for FRODO, due to its treatment of $\beta$ and the $f_i$'s as piecewise constant functions on bins of equal width. When the true densities are exponential, for any group $i$ it is highly probable that most of the $X_{ij}$'s will be near 0, with a few very large measurements in the groups with small rates $\lambda_i$. If the dimensionality (equivalently, the number of bins) $K$ is taken too small, then the groups with large rates will all have estimated $f_i$'s with probability mass near one in the first bin, and mass near zero in the rest. Thus, it is necessary to use a fairly large $K$ in order to capture the differences between these densities. However, this introduces an opposing challenge due to the sparsity of large $X_{ij}$'s: near the right end of the domain, many of the bins will not contain any covariate measurements, so there is little information with which to estimate the densities --- and therefore, the regression function --- in that region. In summary, when the density scales differ to this extent, the ``resolution'' of the data varies throughout the domain.

The use of unequal-width bins would perhaps mitigate this problem, but recall from Section \ref{sec:basis} that the P-spline constructions used here are predicated on an assumption of equally-spaced ``knots'' (which, with splines of degree zero, are simply the bin endpoints). Without these, the unaltered finite-difference penalties on the coefficients no longer serve as approximations to derivatives of a suitable order. It then becomes nontrivial to penalize the $f_i$'s towards some predetermined ``smooth'' shape, although \citet{Li2022} proposed a method of modifying the P-spline penalty in the presence of uneven knots. We do not pursue this here, acknowledging that FRODO in its current state has slightly more difficulty using scale information in the covariate densities than it does using location or shape information.

For this dataset ($N= 200$ groups, each of size $n = 50$), we use parameter values $\left(\alpha, \tilde{\beta}, \sigma_Y\right) = \left(0.1, -0.9, 0.1\right)$. A preliminary visual inspection of KDE's or histograms (not shown) of the covariate data --- and the observation that they are all strictly positive and highly concentrated near zero --- justifies a random walk prior of order $r = 2$ on the densities. In order to capture the ``high-resolution'' differences between covariate measurements near zero as described above, we use a moderately large basis of size $K = 20$. With no reason to suspect severe deviations from this shape we once again set $\delta_i = 0.1$ for all groups. The observed covariates range from $1.3232 \times 10^{-4}$ to 16.3810. Zero is a natural choice for the left endpoint of the assumed domain, and because there are so few large values, we simply take the right endpoint to be the overall sample maximum 16.3810.

\begin{figure}
\centering{\includegraphics[width=\textwidth]{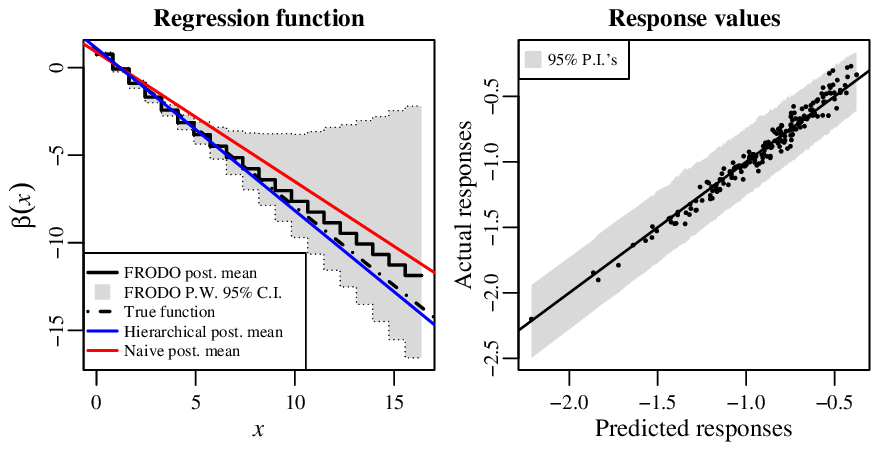}}
\caption{Results of FRODO applied to data with exponential covariates and a linear regression structure. Left: the regression function estimated by FRODO, alongside its pointwise 95\% credible region, the true function, and posterior mean estimates from hierarchical and naive scalar models. Right: responses $\hat{Y}_i$ predicted by FRODO (along with 95\% prediction intervals) vs.\ the true response values.} \label{fig:exp_linear_reg}
\end{figure}

The regression results in the left plot of Figure \ref{fig:exp_linear_reg} represent the most significant example of the phenomenon discussed in Section \ref{sec:normal_quad}; namely, the heightened uncertainty in the regression function in regions where covariate measurements are sparse. Here, 99.73\% of the observed $X_{ij}$'s lie in the left half of the domain, while all of the latent $\lambda_i^{-1}$'s lie within the first 3 bins. Thus, the pointwise 95\% credible interval for $\beta$ is quite narrow near zero --- where most of the covariates are concentrated --- and becomes significantly wider moving from left to right. Once again, we compare FRODO to two scalar models: a naive linear regression using the of the $\bar{X}_{i}$'s as fixed covariates, and a hierarchical linear model in which the latent $\lambda_i$'s are jointly inferred with the regression parameters. Once again, the estimated regression function from the hierarchical model is very close to the true function, and the FRODO estimate approximates it quite well. Some attenuation bias occurs in the right half of the domain, but because all of the covariate densities have such small mass in this region, this does not seem to adversely affect the regression inference in any other significant way. Indeed, the right plot of Figure \ref{fig:exp_linear_reg} shows that the predicted responses closely align with the true $Y_i$'s.

\begin{figure}
\centering\includegraphics[width=\textwidth]{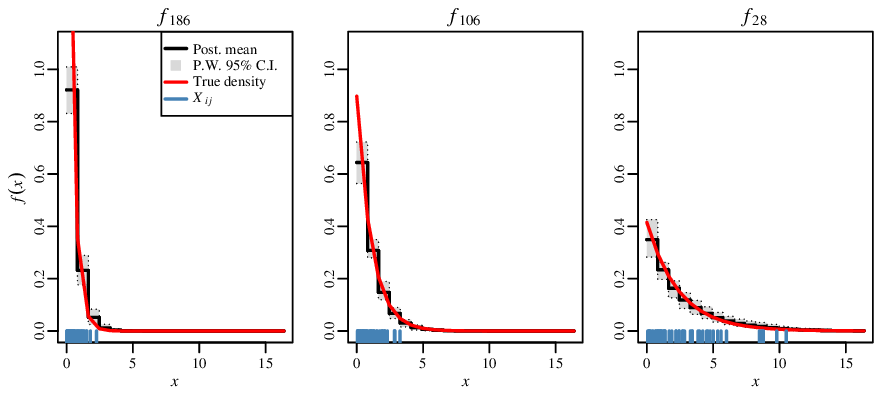}
\caption{For a selection of groups (from the data with exponential covariates and a linear regression structure), the FRODO estimate of the group-specific covariate density, alongside its pointwise 95\% credible region. The true densities are superimposed as red lines, and the actual covariate samples are shown as rug plots.} \label{fig:exp_linear_dens}
\end{figure}

As in previous studies, we compare inferred and true covariate densities for multiple groups in Figure \ref{fig:exp_linear_dens}. FRODO appears to do a good job of capturing the true densities for small, moderate, and large $\lambda_i$'s, although with no real deviations from the shape imposed by the random walk prior, this is perhaps not surprising.

\subsection{Beta covariate densities, linear regression model} \label{sec:beta_linear}
In the following two sections, we demonstrate FRODO's ability to capture regression relationships that are encapsulated in the shapes of the covariate densities, rather than their locations or scales. Whereas the covariate densities in preceding examples were governed by group-level latent parameters which were random themselves, here those parameters are deterministic, allowing us to better control the range of shapes we see. In particular, for this section we take $\xi = \left(\xi_1, \ldots, \xi_N\right)$ to be a mesh of equally-spaced points from $1/10$ to 9/10, and
\begin{align}
X_{ij} \sim \mathrm{Beta}\left(\xi_i, 1 - \xi_i\right). \nonumber
\end{align}
The regression model is
\begin{align}
Y_i &= \alpha + \tilde{\beta}\xi_i + \epsilon_i \nonumber \\
&= \alpha + \mathbb{E}_i\left[\tilde{\beta} X \right] + \epsilon_i. \nonumber
\end{align}
The true densities $f^*_i$ are bimodal for all $i$, with peaks at 0 and 1 and minima at $1/2$. For small $i$ with $\xi_i < 1 - \xi_i$, the peak on the left is wider than the one on the right, so $f_i$ is skewed towards 0 and $\mathbb{E}_i\left[X\right] < 1/2$. The opposite is true for large $i$, and for $i$ near $N/2$ the densities are roughly symmetric.

For this simulation, we use $N = 250$ groups. Because the beta densities have relatively low variance (for the parameter values used here, all of them have variance below 1/8), we use relatively small groups of size $n_i = 15$ for all $i$, so that the difference between the ``true'' and ``naive'' regression functions is more pronounced\footnotemark. The true regression parameters are $\left(\alpha, \tilde{\beta}, \sigma_Y\right) = \left(0.2, 1, 0.05\right)$.
\footnotetext{With large groups, the ``naive'' regression with group-level covariate sample means would be quite close to the true model, making it difficult to tell which one FRODO was capturing.}

Upon inspection of the available covariate data, one would see that all covariate measurements are constrained to the unit interval, with the minimum and maximum measurements being extremely close to 0 and 1, respectively. Thus, $[a',b'] = [0,1]$ is a sensible choice for the assumed domain. Quick visual assessment of KDE or histogram estimates for the group-specific covariate densities reveals that they are neither Gaussian nor exponential. This observation, combined with the strong evidence that the densities are supported only on the unit interval, may lead one to believe that the covariates within each group are, indeed, roughly Beta-distributed. This justifies a random walk prior of order $r = 1$ on the densities, for which the limiting shape is a uniform distribution. Note, however, that unlike the examples above for which we used second- and third-order random walk priors, here the limiting behaviour is \textit{unique}, in the sense that there is only one uniform density on the chosen domain. Thus, if all groups had small smoothing parameter scales $\delta_i$ (corresponding to a prior assumption that no severe deviations from the limiting shape occurred), the FRODO estimates of the covariate densities all would be nearly identical, thereby suppressing the differences between groups and compromising the model's ability to extract meaningful regression information. With an assumed first-order random walk prior, one should therefore expect that the covariate densities will exhibit larger deviations from the limiting shape than they would in a situation where $r > 1$ was appropriate (especially since a bimodal shape will be apparent for at least some of the groups upon preliminary visual inspection). Thus, rather than the default $\delta_i = 0.1$ used in previous examples, here we take $\delta_i = 1$ for all groups. Finally, since several groups have most of their covariate measurements near the endpoints (necessitating bins which are narrow enough to capture differences in densities within these regions), we use $K = 12$ bins: more than the 10 used in the Gaussian examples, but less than the 20 used in Section \ref{sec:exp_linear} since we do not have enough covariate measurements per group to support such a large number of bins (especially since ``roughness'', or deviation from the random walk shape, is penalized less severely here).

\begin{figure}
\centering{\includegraphics[width=\textwidth]{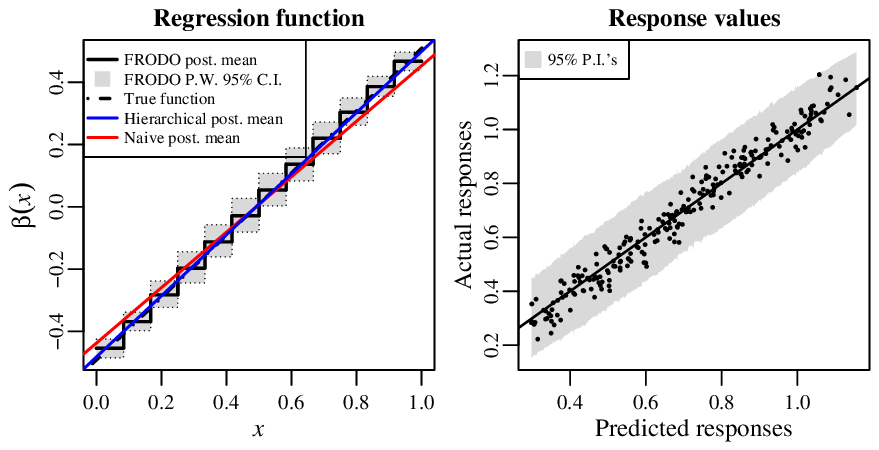}}
\caption{Results of FRODO applied to data with beta-distributed covariates and a linear regression structure. Left: the regression function estimated by FRODO, alongside its pointwise 95\% credible region, the true function, and posterior mean estimates from hierarchical and naive scalar models. Right: responses $\hat{Y}_i$ predicted by FRODO (along with 95\% prediction intervals) vs.\ the true response values.} \label{fig:beta_linear_reg}
\end{figure}

Once again, the regression component of the model is visualized in Figure \ref{fig:beta_linear_reg}, alongside posterior mean estimates from naive and hierarchical scalar models
In contrast to previous datasets, here there are more covariate measurements at each endpoint of the domain than there are in the middle, leading to a slight ``bulge'' in the pointwise 95\% credible interval around 0.5. However, each bin is relatively well-populated with observations, compared to the large differences in concentration seen in previous examples. It is visually obvious that FRODO captures the true regression function and not the naive one. The plot of predicted vs.\ true responses on the right of Figure \ref{fig:beta_linear_reg} provides further confirmation that FRODO's regression inference is satisfactory here.

\begin{figure}
\centering\includegraphics[width=\textwidth]{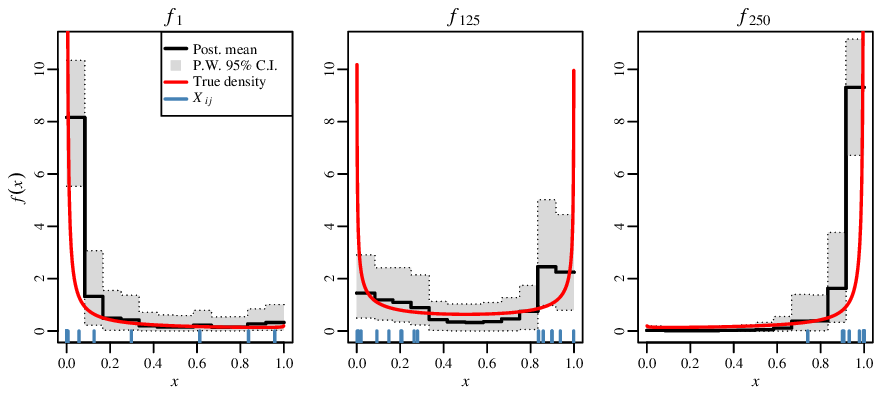}
\caption{For a selection of groups (from the data with beta-distributed covariates and linear regression structure), the FRODO estimate of the group-specific covariate density, alongside its pointwise 95\% credible region. The true densities are superimposed as red lines, and the actual covariate samples are shown as rug plots.} \label{fig:beta_linear_dens}
\end{figure}

Figure \ref{fig:beta_linear_dens} shows that FRODO has more difficulty inferring the true densities here than for previous examples. Although the asymmetrical shapes for $\xi_i$'s near 0.1 or 0.9 are captured, the steep curvature of the true densities near the endpoints in these cases results in them being near the edges of the model's pointwise 95\% credible intervals --- if not excluded altogether --- in these regions. From the middle plot, we see that the model imposes a somewhat excessive degree of uniformity on the nearly-symmetric densities for which $\xi_i$ is near 0.5. These difficulties are not surprising: given the small group sizes and the fairly large values used for $K$ and the $\delta_i$'s, neither the prior nor the likelihood make very strong implications about the density shapes. Aside from collecting more covariate measurements for each group (i.e.\ strengthening the likelihood), the only other possible mitigation for this would be to strengthen the prior: either by using smaller $\delta_i$'s to more strictly enforce the uniform shape, or by using a smaller $K$ to reduce the dimensionality of the problem. However, as discussed above, both of these options would result in an obfuscation of any information that does exist in the available covariate data. Thus, the most prudent choice seems to be accepting that FRODO's density inference in this example is necessarily limited to some degree. Fortunately, this limitation does not adversely affect any of the inference on the regression side of the model. Furthermore, despite the relative ``roughness'' of the FRODO density estimates\footnotemark, they are certainly improvements over, say, ``raw'' histograms (corresponding to $\delta_i \to \infty$), for which the low amount of covariate data would result in even less interpretable shapes.
\footnotetext{Note that this is an inherent difficulty in any dataset for which the first-order random walk prior is justified, because imposing smoothness in this case is inseparable from forcing all of the densities towards being identical.}

\subsection{Beta covariate densities, nonlinear regression model} \label{sec:beta_quad}
Although the previous example shows that FRODO can extract relationships based on the shapes of covariate densities, the regression model itself still ultimately depended only on the means of the covariate measurements. The non-additive structure of the $X_{ij}$'s would pose a challenge for many established multilevel methods, but it is conceivable that one could devise a nonparametric, hierarchical Bayesian method which jointly inferred the $\mathbb{E}_i\left[X\right]$'s while using them to recover the correct regression parameters, subverting the need for full functional regression on the densities. When the regression is not linear, this may not be the case. Thus, in this section we combine a nonadditive covariate structure with a nonlinear regression model to demonstrate the full generality of FRODO. Once again $\xi$ is a mesh of equally-spaced points, this time from $1/10$ to 2, and
\begin{align}
X_{ij} &\sim \textrm{Beta}\left(\xi_i, \xi_i\right), \nonumber \\
Y_i &= \alpha + \tilde{\beta}\left(1+ \frac{1}{2\xi_i + 1}\right) +  \epsilon_i \nonumber \\
&= \alpha + \mathbb{E}_i\left[4\tilde{\beta} \left(X-\frac{1}{2}\right)^2\right] + \epsilon_i .\label{eq:beta_square_reg}
\end{align}
Here, the regression function is $\beta^*(x) = 4\tilde{\beta}(x-1/2)^2$. The $f^*_i$'s are all symmetric: bimodal and U-shaped for $i$ near 1, roughly uniform for $i$ near $N/2$, and peaked at 1/2 for $i$ near $N$. For positive $\tilde{\beta}$, the expected response $\mathbb{E}_i\left[Y\right]$ is higher for ``more bimodal'' covariate densities and lower for ``more unimodal'' ones. The regression is therefore entirely dependent on the shapes of the densities, not their locations or scales. Furthermore, because the densities are all symmetric it holds that $\mathbb{E}^*_i\left[X\right] = 1/2$ for all $i$. Thus, any modelling approach targeting $\beta\left(\mathbb{E}_i\left[X\right]\right)$ (``regression on the expectaton'') will be unsuitable here\footnotemark, as opposed to FRODO with its use of ``the expectation of the regression'', $\mathbb{E}_i\left[\beta\left(X\right)\right]$. In every aspect, this particular data structure is decidedly ``non-classical'', and FRODO seems uniquely well-suited to handle such a structure.
\footnotetext{In theory, one could invoke a measurement error method with more general assumptions on the covariate structure. Recall from \ref{sec:frodo_intro} that the frequentist approach of \citet{Hu2008} described in Section \ref{sec:frodo_intro} assumed a general functional mapping the $f^*_i$'s to the $\xi_i$'s. Although higher-order moments should be permissible under their assumptions, the authors required a \textit{known} functional. Thus, even with their level of generality it would still be necessary to assume quadratic regression \textit{a priori.}}

Because the true covariate densities all have expectation equal to 1/2, the regression function is actually not unique: indeed, when the $f^*_i$'s are all symmetric Beta densities, (\ref{eq:beta_square_reg}) is equivalent to $\alpha + \mathbb{E}\left[4\tilde{\beta} X^2\right] + \epsilon_i$, up to a term which is constant with respect to $i$. This does not seem to be a problem in practice, however: even when HMC chains are explicitly initialized such that $\beta$ is close to the latter form, they converge to a posterior which is consistent with (\ref{eq:beta_square_reg}). We conjecture that the FRODO posterior concentrates around the form of the regression function with ``lowest error'': empirically, we observed that the within-group sample means of $\left(X_{ij} - 1/2\right)^2$ values provide much more accurate estimates of their population analogues than the within-group sample means of the $X_{ij}^2$'s.

For this example, we simulated a dataset with $N = 250$ groups, each containing $n = 60$ covariate measurements. The true regression parameters were $\left(\alpha, \tilde{\beta}, \sigma_Y\right) = \left(0.7, 1, 0.1\right)$. As in Section \ref{sec:beta_linear}, the observed range of the covariate measurements provides strong evidence that $[0,1]$ is a good choice for the assumed density domain. Here, the range of shapes in preliminary histograms or KDE's (from bimodal, to roughly uniform, to unimodal) gives further justification for a random walk prior of order $r = 1$. As in the previous section, we take $\delta_i = 1$ for all $i$ to allow a greater degree of deviation from the limiting (uniform) shape of the prior. Because the data is highly concentrated near the endpoints for the groups whose $\xi$-values are low (even moreso than in Section \ref{sec:beta_linear}'s dataset), we use a basis of size $K = 15$.

Due to the aforementioned uselessness of methods involving ``regression on expectations'' here, constructing scalar models to compare with FRODO is nontrivial. We cannot use a ``naive GAM'' as we did for the Gaussian quadratic model in Section \ref{sec:normal_quad}. There, $\mathbb{E}_i\left[X^2\right]$ and $\left(\mathbb{E}_i\left[X\right]\right)^2$ differed by a constant, but this is not the case here. Thus, the naive scalar model we use for comparisons is somewhat contrived: a linear regression model, using the within-group sample means of the $\left(X_{ij} - 1/2\right)^2$ values as covariates. As always, the hierarchical scalar model assumes the true forms of the regression function and covariate densities are all known, jointly inferring the $\xi_i$'s and all regression parameters.

\begin{figure}
\centering{\includegraphics[width=\textwidth]{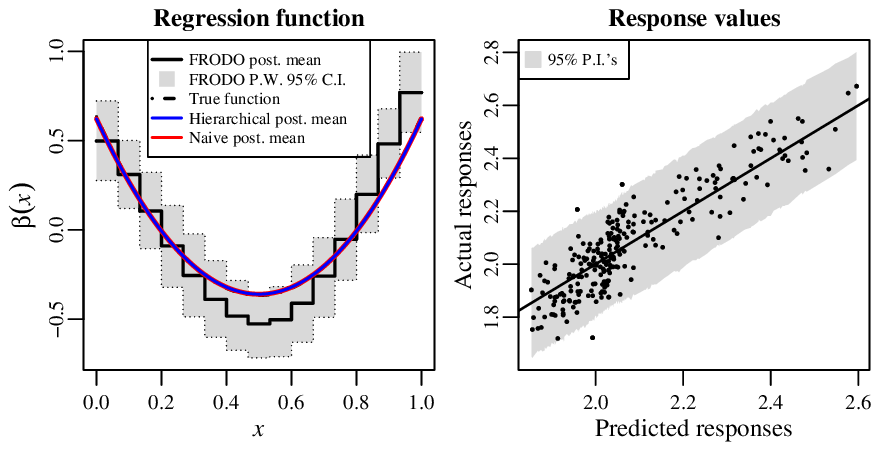}}
\caption{Results of FRODO applied to data with beta-distributed covariate data and a quadratic regression structure. Left: the regression function estimated by FRODO, alongside its pointwise 95\% credible region, the true function, and posterior mean estimates from hierarchical and naive scalar models. Right: responses $\hat{Y}_i$ predicted by FRODO (along with 95\% prediction intervals) vs.\ the true response values.} \label{fig:beta_square_reg}
\end{figure}

Because of the relatively large group sizes, and the fact that the quadratic form of the regression function was assumed known in both scalar models, the naive model does not suffer from any appreciable attenuation bias. As shown on the left of Figure \ref{fig:beta_square_reg}, both it and the hierarchical scalar model approximate the true regression function almost perfectly. Some bias is apparent in the FRODO estimate, particularly near the vertex at 1/2, but its pointwise 95\% credible interval almost completely captures the true function. On the right side of Figure \ref{fig:beta_square_reg}, we see a moderate ``clumping'' of predicted responses just over 2.0, where the variability in the actual $Y_i$'s exceeds that of the mean predictions from FRODO. These values correspond to groups with $\xi$-values near 1 (i.e.\ those whose true covariate densities $f^*_i$ are close to uniform). For this dataset, it appears that FRODO has a small amount of difficulty capturing small shape differences between nearly-uniform densities. Note also that a few groups have posterior $95\%$ prediction intervals which exclude their observed responses, although it seems reasonable to attribute this to mere random chance given the large number of groups. In any case, the overall fit appears largely satisfactory, especially considering that the true forms of the regression function and covariate densities are not known \textit{a priori}.

\begin{figure}
\centering\includegraphics[width=\textwidth]{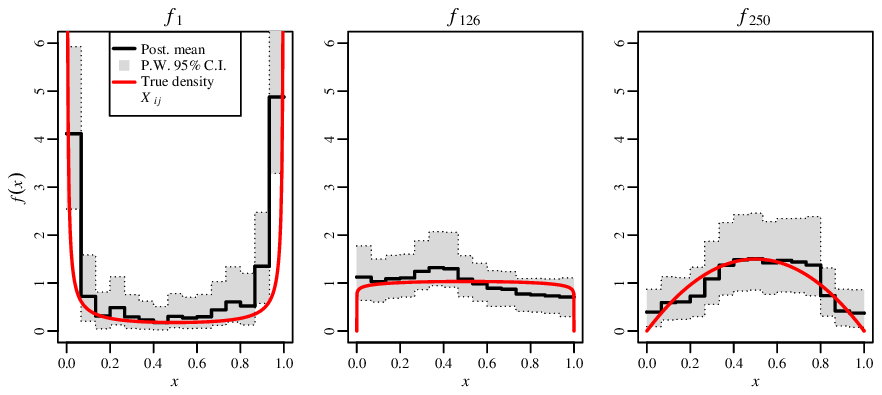}
\caption{For a selection of groups (from the data with beta-distributed covariates and quadratic regression structure), the FRODO estimate of the group-specific covariate density, alongside its pointwise 95\% credible region. The true densities are superimposed as red lines, and the actual covariate samples are shown as rug plots.} \label{fig:beta_square_dens}
\end{figure}

Figure \ref{fig:beta_square_dens} shows that FRODO roughly captures all three types of density shapes present in this data, although some excess noise and bias is evident in the posterior estimates. This is particularly evident for the unimodal density in the right plot. Although the true density is fully contained in the pointwise 95\% credible interval, the posterior mean is perhaps somewhat too flat. The true unimodal densities in this dataset certainly differ more subtly from the uniform shape than the bimodal ones (contrast the true density in the left plot of Figure \ref{fig:beta_square_dens} with that on the right) --- since the prior on densities here is structured only in terms of ``deviations from uniformity'', this slight deficiency is not entirely unexpected. As in Section \ref{sec:beta_linear}, some of the excess noise in the density inference is an unavoidable consequence of the larger values of $K$ and $\delta$ necessary to capture the shapes and fine structure of the true densities with the first-order prior.

\section{Extended simulation study: FRODO with varying group sizes and a group-level covariate} \label{sec:croon}
As a final ``application'' of FRODO, we recreate the simulated data considered by \citet{Croon2007}. This is very much a ``classical'' model, with Gaussian covariate data and a linear regression function much like the one considered in Section \ref{sec:gauss_lin}. However, there are three unique features here which were absent from the ``toy'' examples explored above. First (recalling the notation of (\ref{eq:covar}--\ref{eq:err})), the parameter values are $\left(\sigma_\xi, \sigma_X, \alpha, \tilde{\beta}, \sigma_Y\right) = \left(1, 3, 0.3, 0.3, \sqrt{0.35}\right)$: not only is the within-group variability of the $X_{ij}$'s much greater than the between-group variability of the true $\xi_i$'s, but the regression error is also quite high, accounting for just under 65\% of the variability in the $Y_i$'s. Overall, the amount of ``signal'' in the data --- at both the covariate and regression levels --- is low relative to the amount of noise. Second, there are varying group sizes, some of which are quite small: out of $N = 100$ groups, roughly 50\% (randomly selected with probability 1/2) contain $n_i = 10$ covariate measurements, and the rest contain $n_i = 40$. Finally, the actual regression model is altered from the basic FRODO form considered thus far, with the inclusion of a ``scalar'' group-level covariate $Z$ as in (\ref{eq:simple_reg}):
\begin{align}
Y_i = \alpha + \tilde{\beta}\xi_i + \beta_Z Z_i + \epsilon_i. \label{eq:croon_reg}
\end{align}
The covariate values $Z_i$ are generated from a standard Normal distribution, independently of $\xi$, and are treated as fixed observations.

It is straightforward to extend FRODO to accommodate for $Z$ by putting a $\mathcal{N}\left(0, 20\sigma_Y\right)$ prior on $\beta_Z$, conditionally independent from the prior for $\beta$ (which still denotes the regression function corresponding to the group-specific densities of the $X_{ij}$'s). We use a third-order random walk prior on the $f_i$'s with $K = 10$ bins as in Section \ref{sec:gauss_lin}, since the available data gives no reason to suspect that finer structures need to be captured. Due to the relatively small amount of covariate measurements, we simply take the assumed domain $[a', b']$ to be the range of observed $X_{ij}$-values, which in this case is $[-12.0365, 11.2258]$. For the groups of size $n_i = 40$, the default smoothing prior scale choice $\delta_i$ is appropriate, but with only $n_i = 10$ observations in the smaller groups, a tighter prior is necessary to ensure posterior density estimates with useful shape information. Thus, we set $\delta_i = 0.05$ for the small groups.

\begin{figure}
\centering{\includegraphics[width=\textwidth]{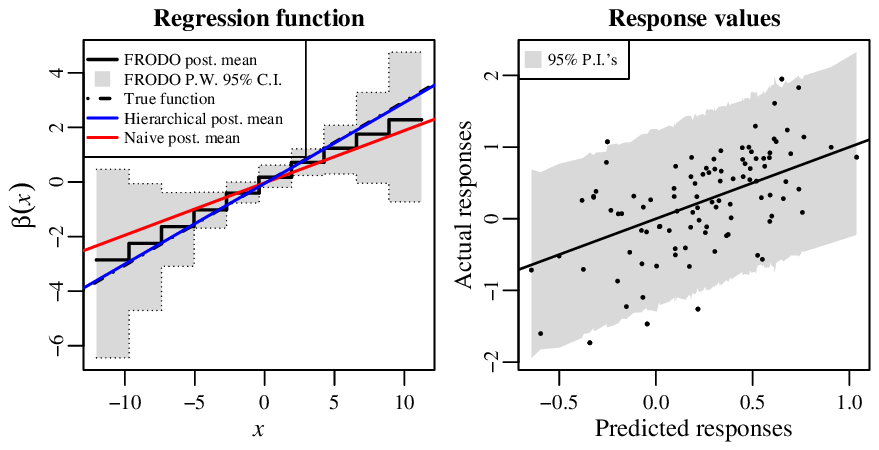}}
\caption{Results of FRODO applied to data with Gaussian covariates, a linear regression structure, and an additional group-level scalar covariate. Left: the regression function for the multilevel covariate estimated by FRODO, alongside its pointwise 95\% credible region, the true function, and posterior mean estimates from hierarchical and naive scalar models. Right: responses $\hat{Y}_i$ predicted by FRODO (along with 95\% prediction intervals) vs.\ the true response values.} \label{fig:croon_reg}
\end{figure}

The actual method proposed by \citet{Croon2007} for micro-macro modelling is frequentist and involves a stepwise estimation procedure. An R implementation exists \citep{micromacro}, but here we are only interested in comparing FRODO to analogous scalar Bayesian methods. Thus, as in the studies of Section \ref{sec:frodo_sim} we compare it to both a naive and hierarchical scalar model, trivially extended to accommodate $Z$ and place a prior on $\beta_Z$. These results are shown in the left plot of \ref{fig:croon_reg}. Note how much wider the pointwise 95\% credible interval is --- particularly near the endpoints --- than the one in the similar model of Figure \ref{fig:norm_linear_reg}, owing to the higher noise and smaller amount of available covariate data here. It appears that the posterior for FRODO has concentrated somewhere in between the true and naive regressions. Indeed, FRODO's posterior mean for $\sigma_Y$ is 0.5987 (95\% C.I.\ $(0.5182, 0.6972)$), in contrast with 0.5857 from the hierarchical scalar model (95\% C.I.\ $(0.5053, 0.6834)$) and 0.6130 from the naive scalar model (95\% C.I.\ $(0.5370, 0.70234)$). Given that the dataset is fairly small and high in noise, it is perhaps unsurprising that FRODO struggles more than it did in previous studies. However, this seems to be a problem of variability, not of bias: other simulated datasets with the exact same parameters, group sizes, and number of groups resulted in FRODO estimates with differing amounts of attenuation (not shown). Even the scalar hierarchical model proved quite variable with other datasets, as its estimate of the regression function did not always align as closely with the true function as it does here. Although the high degree of noise in the right plot of Figure \ref{fig:croon_reg} may appear troubling, this is reflective of the actual amount of noise in the data: a plot of predicted vs.\ actual responses from a frequentist multiple linear regression using the true $\xi_i$'s appears similar.  

\begin{figure}
\centering\includegraphics[width=\textwidth]{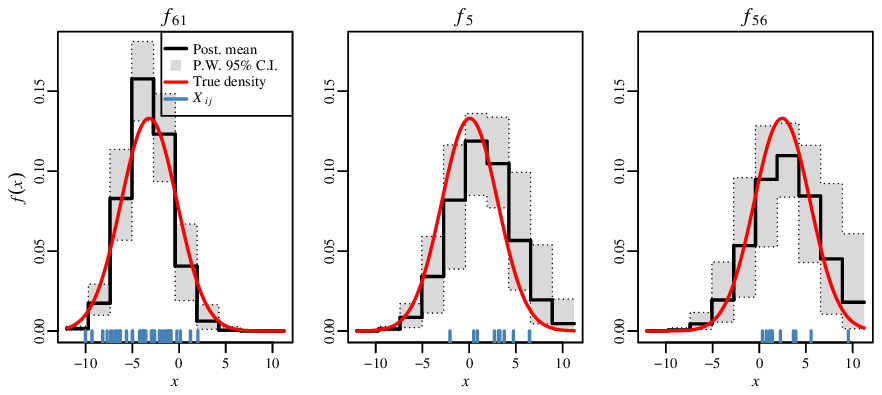}
\caption{For a selection of groups (from the data with Gaussian covariate data, a linear regression structure, and an additional group-level covariate), the FRODO estimate of the group-specific covariate density, alongside its pointwise 95\% credible region. The true densities are superimposed as red lines, and the actual covariate samples are shown as rug plots.} \label{fig:croon_dens}
\end{figure}

The usual density plots are shown in Figure \ref{fig:croon_dens}. Note that the group in the left plot contains 40 individuals, and the other two contain only 10. It is intuitive that the smaller groups would have wider pointwise credible intervals for their densities (on further inspection, this pattern also seemed to hold for other groups not shown here), although it is somewhat noteworthy that the smaller $\delta_i$-values for these groups do not seem to neutralize this effect. Some bias in the model is evident, particularly in the middle plot, but overall the inference provided by FRODO seems reasonable.

\section{Discussion}
In this paper, we have presented a new approach for micro-macro modelling which combines density estimation and functional data analysis into a unified hierarchical Bayesian framework. Although FRODO is relatively simple in principle due to its use of step functions and only \textit{linear} functional regression terms, it is deceptively powerful in its ability to use these elements for approximation of generalized additive models. Beyond the generality of the regression component of the model, FRODO is also quite flexible in terms of the individual-level covariate structures it can accommodate. Whereas many Bayesian methods for GAM's with measurement error or micro-macro structure assume a Gaussian --- or at the very least, additive --- error structure in the $X_{ij}$'s, FRODO has no such limitation, allowing for covariate densities which influence the group-level regression responses through their locations, scales, or shapes. All that is required is the selection of a suitable prior structure for the densities, based on either prior domain knowledge, or --- if this is not possible and an empirical Bayesian approach is required --- a preliminary heuristic examination of the data. Although FRODO's inference on the covariate densities is generally more accurate when the true densities adhere to the specified ``smooth shape'' encoded in the prior, this is not a strict \textit{requirement} provided hyperparameters are chosen carefully.

The simulation studies conducted above show that the power and generality of FRODO translate from theory to practice, providing reasonable inference for a variety of data structures. However, the potential for improvements and extensions to the model is vast. The most immediate potential for this is in the density part of the model, as described in Section \ref{sec:dens_model}. Here we have not considered $r^\mathrm{th}$-order random walk priors for any integer $r > 3$. These would result in densities being penalized towards exponentiated polynomials of higher degree: with an $r^\mathrm{th}$-order random walk prior, $\log f_i (x)$ is close to a polynomial of degree ${r-1}$ when the smoothing parameter $\tau_i$ is small. Such limiting smooth shapes correspond to \textit{generalized error distributions} \citep{Varanasi1989} (or folded versions thereof) with shape parameter $r-1$, of which the Normal, Laplace, and uniform distributions are special cases. For $r > 2$, the generalized error distribution has lighter tails than a Gaussian. It is not certain how useful such higher-order random walk priors would be in practice (i.e.\ how often one might expect covariate densities to be similar to, say, an exponentiated quartic), but one challenge in implementing these would be determining suitable distributions for the ``free parameters'' $\theta_{ik}$, $2 \leq k \leq r$. Equivalent derivations of the type carried out for $r = 2$ and 3 in Section \ref{sec:dens_model} would be much more complex.

There is even room for generalization within the confines of the third-order (resp.\ second-order) random walk priors considered here. Although the construction in Section \ref{sec:dens_model} was explicitly tailored in terms of Gaussian (resp.\ exponential) distributions, in principle it could be adapted for \textit{any} densities whose logarithms are roughly quadratic (resp.\ linear) in shape. Folded or truncated Normal distributions may be a useful shape to accommodate with a third-order random walk prior; one could even modify it to allow for densities $f$ such that $\log f$ is approximately quadratic with \textit{positive} leading coefficient, not negative as for a Gaussian. This may be useful for modelling ``U-shaped'' densities, such as the Beta distributions considered in Section \ref{sec:beta_linear}. Similarly, the second-order structure could be generalized to allow for positively-sloped densities (i.e.\ ``reversed'' exponentials), or Laplace densities whose logarithms are \textit{piecewise} linear. Furthermore, it may be useful to combine differing random walk orders within the same model. For instance, the example in Section \ref{sec:beta_quad} might have benefited if we used a third-order random walk prior for the unimodal densities (since symmetric Beta densities are close to Gaussians in shape for large parameter values), a first-order R.W.\ prior for the flatter densities, and perhaps an ``inverted'' third-order R.W.\ prior for the U-shaped densities as suggested above.

Further investigation of the relationships between $n$, $r$, $K$, and $\delta$ would also be useful, particularly how best to set the latter two in terms of the former two. Although the empirical heuristic methods employed here worked well in practice, a more formal approach might result in better performance and generalization. Appeals to asymptotics could guide derivation of mathematical relationships between the hyperparameters: for instance, an expression for an ``optimal'' $\delta_i$ in terms of $r$, $K$, and $n_i$, based on the ``big-O'' relationships shown by \citet{Silverman1982} to guarantee convergence of penalized density estimators in the frequentist setting. The choice of the assumed domain for the densities may also have an effect on any such expressions.

On the subject of ``big-O'' considerations, recall that the difference between the ``true'' and ``proxy'' models ((\ref{eq:frodo_int}) and (\ref{eq:frodo_int_basis}), respectively) was swept under the rug, with a passing acknowledgement that a \textit{truly} formally valid treatment would consider a combined error (\ref{eq:combined_error}) accounting for the use of finite-dimensional approximation in this nonparametric setting. Further study of this approximation error -- including its large-sample behaviour -- would be of interest.

There is also significant potential for generalizations on the regression side of the model. The most immediate of these is the realization of our proposed extension to non-Gaussian responses such as count or categorical data. Just as the regression part of FRODO for the Gaussian responses considered here is nothing more than a functional linear model, allowing for other response types is simply a matter of using functional GLM machinery.

Perhaps the most useful immediate extension to FRODO would be the incorporation of multiple multilevel covariates. Indeed, many real-world micro-macro datasets include several covariates measured at the individual level within groups \citep[e.g][]{Croon2007, Amendola2020,Daouda2021}. Of course, this would increase the computational complexity of FRODO, as the number of parameters to infer grows roughly linearly in the number of multilevel covariates. Note, however, that real-world micro-macro datasets commonly include ordinal covariates with a small number of levels \citep[e.g.][]{Amendola2020, Daouda2021}. Modelling the distributions for these covariates requires only as many basis functions as there are levels, which would mitigate computational difficulty to some extent in practice.

A powerful yet challenging improvement would be modelling more complex relationships amongst covariates. For instance, \citet{Croon2007} considered a version of the simulation study replicated in Section \ref{sec:croon} where the latent and observed group-level covariates ($\xi$ and $Z$, respectively) were correlated \citep[see also measurement error literature such as][]{Richardson1993}. Accounting for dependence between multilevel and ``scalar'' covariates in FRODO will be highly nontrivial, especially if one wishes to maintain flexibility in the shapes of the inferred densities. For instance, if the multilevel data is Gaussian as in Section \ref{sec:croon}, the most obvious way to account for correlation between $\xi$ and $Z$ is to explicitly include it in the prior for the $\xi_i$'s (see Section \ref{sec:dens_model}). However, we have found in practice that the $\xi_i$'s inferred by FRODO are often poor approximations for the actual latent group means of the $X_{ij}$'s, unless a Gaussian shape is heavily enforced on the $f_i$'s by deliberately taking very small $\delta_i$'s. This was not a problem for the examples in Sections \ref{sec:gauss_lin} and \ref{sec:normal_quad}, as the posterior density estimates ended up being close enough to the true Gaussians that there were no major difficulties in the inference. If such latent density parameters are required more explicitly to model correlations with scalar covariates, this inaccuracy may become problematic. The potential for dependence between distinct \textit{multilevel} covariates is arguably even more interesting. Presumably this would require regression on multiple integrals over their joint densities. However, even with the degree-zero splines considered here, this would result in a substantial increase in computational complexity. Indeed, the number of coefficients required to model the joint density of $d$ multilevel covariates for a single group in this way is exponential in $d$. Therefore, some type of simplification would likely be required to make interactions between multilevel covariates viable. See \citet{Lambert2006} for a discussion of multivariate density estimation with splines in the case of a single density.

We conclude by acknowledging potential shortcomings in FRODO for which there are likely no solutions, either due to the inherent properties of the model or the excessive computational difficulty that would be required to solve them. First, one may question the use of piecewise constant basis functions, since higher-order splines would certainly result in smoother and better-behaved density estimates. However, recall from Section \ref{sec:reg} that this choice was made partially for computational convenience: it ensures that the integral of $\beta \cdot f_i$ is simply the inner product of the two functions' coefficients. This is no longer the case with higher-order splines, for which the integrals are more complicated expressions involving products between neighbouring coefficients. Beyond the heightened complexity, we also found in preliminary experiments that the resulting posterior geometry was extremely difficult to navigate with NUTS. Note that these experiments modelled the densities themselves with higher-degree splines, requiring (among other things) a potentially costly softmax transformation of each $\theta_i$ vector. The other possibility is modelling the \textit{logarithms} of the densities with splines \citep[e.g.][]{OSullivan1988}. These approaches are equivalent for degree-zero splines, but with higher degrees the logarithmic approach requires approximate numerical integration to normalize the $f_i$'s, which are exponentiated piecewise polynomials. These numerical integrals, in turn, depend on the spline coefficients in complex ways which would likely complicate the posterior geometry even further. Thus, unless a radically different approach is used to fit the model, higher-order splines do not seem to be worth the effort, given the satisfactory results obtained with piecewise constant functions and the prevalence of ordinal covariates in real-world micro-macro data.

In earlier experiments (not shown), we found problems with bias and sampling efficiency when the within-group covariate noise was large relative to either the regression noise or between-group covariate scale. In the notation of the Gaussian model, problems occurred when the $n_i$'s were small and $\sigma_X$ was large relative to either $\sigma_\xi$ or $\sigma_Y$, especially when the magnitude of the effect size $\tilde{\beta}$ was large. This problem also affected hierarchical scalar models --- suggesting that there is innate difficulty in the posteriors induced by such datasets --- but FRODO did seem slightly more sensitive to it, in the sense that some parameter combinations were problematic for FRODO but not for a scalar model. These problems could be mitigated with different prior choices such as a zero-avoiding prior for $\sigma_Y$, but these can create bias \citep{Gelman2006}. Fortunately, we suspect that the relative noise levels which tend to create problems are unlikely to occur in practice, as they imply either extremely low-error regression models or high-error covariate groups.

Finally, it bears repeating that FRODO only models responses in terms of \textit{expectations of functions of covariates}: any regression relationship that cannot be expressed in the form (\ref{eq:frod_gam}), or some multivariate extension thereof, is incompatible with this methodology. In particular, responses which depend on the medians or modes of densities cannot be modelled with FRODO, requiring other methods specifically suited for those purposes \citep[e.g.][]{Hu2008}. Its current inability to model \textit{functions of expectations} may also be a shortcoming. For instance, if the data in Section \ref{sec:normal_quad} was modified so that the covariate densities had unequal variances and the group-level responses were proportional to these variances, FRODO would not be usable due to the nonconstant $\left(\mathbb{E}_i\left[X\right]\right)^2$ term in the regression. One could potentially augment (\ref{eq:frod_gam}) with an ``outer function'', using terms of the form $g\left(\mathbb{E}_i\left[\beta\left(X\right)\right]\right)$ with some unknown function $g$ to be modelled with a basis function expansion. However, this would likely create a litany of problems with unidentifiability.

Despite these challenges, we believe that FRODO's power and flexibility make it a strong addition to the field of micro-macro regression modelling, especially as improvements and extensions are developed to handle an even broader variety of data structures.



%
\section*{Acknowledgements}
Shaun McDonald wishes to thank the defense committee who reviewed the Ph.D.\ thesis in which this work originally appeared.

\subsection*{Declarations}
This manuscript is largely an adaptation of work originally included in Shaun McDonald's doctoral thesis \citep{mcdonald2022}.
\subsection*{Conflict of interest} The authors have no conflicts of interest to report.
\subsection*{Funding} During the time of this work, Shaun McDonald was affiliated with Simon Fraser University and supported by an NSERC Alexander Graham Bell Canada Graduate Scholarship.  Alexandre Leblanc, Saman Muthukumarana, and David Campbell are also supported by NSERC.

\begin{appendices}
\section{Details of the implementation of FRODO in Stan} \label{chapter:app_frodo}
This appendix expands on the brief discussion in Section \ref{sec:imp} regarding the Stan implementation of FRODO. We explain our method of initializing HMC chains, detail the parameter values used in the NUTS sampler, and assess the sampling behaviour of the simulation studies in Sections \ref{sec:frodo_sim}--\ref{sec:croon}. The reader may also refer to our source code at \url{https://github.com/ShaunMcDonald1021/FRODO}.

This appendix will assume the reader is familiar with Stan, and the terminology associated with implementation and assessment of models therein. However, references to relevant Stan documentation are included where appropriate.

\subsection{Reparameterizations} \label{sec:reparam}
It is known that Stan's sampling behaviour can suffer in the presence of difficult posterior geometries: for instance, when the posterior has heavy tails or nonlinear correlations between parameters \citep[][Section 25.7 of the User's Guide and references therein]{stanref}. Following standard advice [ibid.], we use \textit{non-centered parameterizations} for various parameters. Briefly, this means restating the target distribution (i.e.\ the posterior) in terms of parameters which do not have the same hierarchical dependence structures as in the original parameterization, thereby inducing a posterior geometry more amenable to HMC. The parameters of interest (see Sections \ref{sec:dens_model}--\ref{sec:reg}) are then recovered as deterministic functions of the ones actually sampled. Additionally, the error variance $\sigma_Y$ is expressed as the ratio of a half-normal random variable and a Gamma random variable with shape parameter 2, neither of which have the type of heavy tails which are often problematic in NUTS \citep{stanref}. The full details of the reparameterizations used are described in the comments of the source code referenced above.

\subsection{Initialization of chains}
By default, Stan initializes all parameters uniformly in the range $[-2, 2]$ (for positive parameters, this is done on the logarithmic scale) \citep{Carpenter2017}. This proved to be a problem for the densities: the default scheme, in conjunction with the reparameterizations discussed in Section \ref{sec:reparam}, almost always resulted in initial density estimates for which the logarithm of the posterior was infinite. It is not known how often these were ``genuine'' infinities as opposed to mere numerical overflow, but in either case the result is an inability to obtain posterior samples.

The problem appears to be related to the random walk structure of the $\theta_i$'s, which are encoded into the Stan model through a linear transformation of ``non-centered'' parameters. This transformation tends to ``magnify'' the variability in the default initial values to the extent that the initial $\phi_i$'s are severely mismatched with the likelihood of their corresponding covariate data (see Section \ref{sec:dens_model}). Thus, we use a modified initialization strategy based on preliminary frequentist estimates for the $f_i$'s. These are obtained using P-splines and Poisson regression models for the bin counts in each group, as proposed by \citet[][Section 8]{Eilers1996}. These are then ``inverse-transformed'' to obtain initial values for the parameterization used in Stan. A modest amount of randomness --- Gaussian noise for the $\theta_i$'s, and Gamma-distributed initial values for the $\tau_i$'s and scale components for the ``free parameter'' means defined in Section \ref{sec:dens_model} --- is injected into the initialization to ensure that the starting points of the HMC chains are reasonably diffuse \citep{Gelman1992}.

\subsection{Parameters of NUTS samplers} \label{sec:samp_param}
Sampling in Stan depends on several ``parameters\footnotemark'' which govern the behaviour of the NUTS algorithm. Section 15.2 of the Stan Reference Manual \citep{stanref} explains these parameters, and further details on their implications for sampling performance are discussed in the vignette at \url{https://mc-stan.org/misc/warnings.html}.
\footnotetext{Not to be confused with the ``parameters'' whose posterior is the target of inference. In Section \ref{sec:samp_param}, the word ``parameters'' refers only to the ``\textit{sampling} parameters'' discussed therein.}

Due to the complexity of FRODO's posterior geometry, we found it necessary to use maximum tree depths and target Metropolis acceptance rates which were higher than the defaults (10 and 0.8, respectively). In all of the simulation studies shown in Sections \ref{sec:frodo_sim}--\ref{sec:croon}, we used a maximum tree depth of 12. The target acceptance rate was set to 0.99, except in the studies with Gaussian covariate data, where it was set to 0.985. For each study, we ran four NUTS chains in parallel. Each chain was run for 750 warmup iterations, then 1250 sampling iterations.

\subsection{Behaviour of simulation runs}
In Table \ref{tab:frodo}, we summarize the performance of the samplers for each of the six simulation studies in Section \ref{sec:frodo_sim}. 
Each study is denoted by the section in which it appears, and the following information is included for each one.
\begin{enumerate}
    \item The maximum warmup time (in seconds) for any of the four chains,
    \item the maximum sampling time (in seconds) for any of the four chains,
    \item the smallest estimated \citep{Vehtari2021} effective sample size ($n_\mathrm{Eff}$) for any parameter in the model, and
    \item the maximum split $\hat{R}$ value for any parameter in the model \citep{Vehtari2021}.
\end{enumerate}
Note that the reported $n_\mathrm{Eff}$ (resp.\ $\hat{R}$) is the minimum (resp.\ maximum) over the actual sampled parameters \textit{and} the ``true'' model parameters obtained with transformations (see Section \ref{sec:reparam}). All simulations were run on an Acer laptop with 16 GB of RAM and four Intel i5-9300H 2.40GHz CPU cores.

\begin{table}
\centering
\begin{tabular}{r|c|c|c|c}
Study & Max.\ warmup time & Max.\ sampling time & Min.\ $n_\mathrm{Eff}$ & Max.\ $\hat{R}$ \\ \hline
\ref{sec:gauss_lin} & 654.280 &  486.304 & 572.618 & 1.004  \\
\ref{sec:normal_quad} & 1350.34 & 2168.11 & 996.3261 & 1.007 \\
\ref{sec:exp_linear} & 719.058 & 1108.260 & 1036.961 & 1.004 \\
\ref{sec:beta_linear} & 447.584 & 831.499 & 857.6613 & 1.007 \\
\ref{sec:beta_quad} & 509.821 & 702.896 & 1022.321 & 1.006 \\
\ref{sec:croon} & 99.554 & 89.056 & 622.911 & 1.009
\end{tabular}
\caption{Various quantities quantifying the performance and sampling behaviour of FRODO, for each of the simulated datasets in Section \ref{sec:frodo_sim}.} \label{tab:frodo}
\end{table}

In every simulation study, all parameters had effective sample sizes exceeding 450. \citet{Vehtari2021} recommend a threshold of at least 550 effective samples per parameter, so we are confident that ours are large enough for inference to be reasonably accurate. \citet{Vehtari2021} recommends considering split $\hat{R}$ values above a threshold of 1.01 to be indicative of convergence problems, and no values in our studies exceeded this threshold.

As one would expect given FRODO's complexity, warmup and sampling are several times slower than they are for the corresponding scalar models used in the simulation studies (not shown). The only study whose computation time we would consider problematic is the one from Section \ref{sec:normal_quad}, with Gaussian covariate data and a quadratic regression structure. Including warmup and sampling, the Stan model for this study took over an hour to run. Most of the sampling iterations for this study had larger tree depths than in the other studies, meaning that the number of gradient evaluations involved in sampling was roughly higher by a factor of 2 or more \citep[][Section 15.2 of Reference Manual]{stanref}. This is likely a consequence of posterior geometry, and the way in which the samplers adapt to it during warmup. However, it should be noted that we deliberately used a liberal number of warmup iterations, and chains appeared to have converged to the ``typical set'' \citep{Betancourt2018} well before sampling began (not shown). Note also that the smallest effective sample size is over twice as large as the threshold of 400 recommended by \citet{Vehtari2021}
Therefore, reasonable posterior inference with acceptable computation time could likely be achieved by reducing the number of warmup and sampling iterations, provided the latter did not induce problematic $\hat{R}$ values.

Finally, recall from Section \ref{sec:frodo_sim} that estimates of the regression variance, $\sigma_Y$, are biased upward in ``naive'' regression models, and this fact can be used to check whether or not FRODO is recovering ``true'' regression relationships. For each simulation study, Table \ref{tab:sigma_y} shows the true value of $\sigma_Y$, as well as the posterior mean and 95\% credible interval for this parameter from FRODO, the hierarchical scalar model, and the naive scalar model (see the beginning of Section \ref{sec:frodo_sim}). The endpoints posterior intervals are simply $0.025$- and $0.975$-quantiles from the HMC samples. For almost every simulation study, the FRODO estimate for $\sigma_Y$ is much closer to the true value than the estimate from the naive model. The sole exception is the example of Section \ref{sec:beta_quad}, with Beta-distributed covariates and a nonlinear regression structure, for which all models produce accurate estimates of $\sigma_Y$. Recall, however, that the large group sizes in this example rendered bias in the naive model negligible.

\begin{table}
\centering
\begin{tabular}{r|c|c|c|c}
Study & True & FRODO & Hierarchical scalar model & Naive scalar model\\ \hline
\ref{sec:gauss_lin} & 0.5 & 0.493 (0.442, 0.547) & 0.490 (0.438, 0.545) & 0.555 (0.511, 0.603)  \\
\ref{sec:normal_quad} & 0.5 & 0.473 (0.385, 0.564) & 0.479 (0.416, 0.550) & 0.886 (0.815, 0.967) \\
\ref{sec:exp_linear} & 0.1 & 0.100 (0.073, 0.128) & 0.097 (0.070, 0.123) & 0.165 (0.150, 0.182) \\
\ref{sec:beta_linear} & 0.05 & 0.065 (0.054, 0.077) & 0.055 (0.046, 0.065) & 0.089 (0.082, 0.098) \\
\ref{sec:beta_quad} & 0.1 & 0.094 (0.084, 0.105) & 0.099 (0.090, 0.109) & 0.101 (0.093, 0.111) \\
\ref{sec:croon} & 0.592 & 0.599 (0.518, 0.697) & 0.586 (0.505, 0.684) & 0.613 (0.537, 0.704)
\end{tabular}
\caption{Posterior inference for $\sigma_Y$ (the regression error) from FRODO and the scalar models within each simulation study. For each model, the posterior mean is reported, as is a 95\% credible interval in parentheses. The second column from the left shows the true $\sigma_Y$.} \label{tab:sigma_y}
\end{table}

\end{appendices}

\bibliographystyle{plainnat}
\bibliography{frodo_bib}

\begin{thebibliography}{44}
\providecommand{\natexlab}[1]{#1}
\providecommand{\url}[1]{\texttt{#1}}
\expandafter\ifx\csname urlstyle\endcsname\relax
  \providecommand{\doi}[1]{doi: #1}\else
  \providecommand{\doi}{doi: \begingroup \urlstyle{rm}\Url}\fi

\bibitem[Aitchison and Shen(1980)]{Aitchison1980}
J.~Aitchison and S.~M. Shen.
\newblock Logistic-normal distributions: Some properties and uses.
\newblock \emph{Biometrika}, 67\penalty0 (2):\penalty0 261--272, 1980.
\newblock \doi{10.2307/2335470}.

\bibitem[Amendola et~al.(2020)Amendola, Barra, and Zotti]{Amendola2020}
Adalgiso Amendola, Cristian Barra, and Roberto Zotti.
\newblock {Does graduate human capital production increase local economic
  development? An instrumental variable approach}.
\newblock \emph{Journal of Regional Science}, 60\penalty0 (5):\penalty0
  959--994, 2020.
\newblock \doi{10.1111/jors.12490}.

\bibitem[Bennink et~al.(2013)Bennink, Croon, and Vermunt]{Bennink2013}
Margot Bennink, Marcel~A. Croon, and Jeroen~K. Vermunt.
\newblock {Micro-Macro Multilevel Analysis for Discrete Data: A Latent Variable
  Approach and an Application on Personal Network Data}.
\newblock \emph{Sociological Methods \& Research}, 42\penalty0 (4):\penalty0
  431--457, 2013.
\newblock \doi{10.1177/0049124113500479}.

\bibitem[Bennink et~al.(2016)Bennink, Croon, Kroon, and Vermunt]{Bennink2016}
Margot Bennink, Marcel~A. Croon, Brigitte Kroon, and Jeroen~K. Vermunt.
\newblock {Micro-macro multilevel latent class models with multiple discrete
  individual-level variables}.
\newblock \emph{Advances in Data Analysis and Classification}, 10:\penalty0
  139--154, 2016.
\newblock \doi{10.1007/s11634-016-0234-1}.

\bibitem[Betancourt(2017{\natexlab{a}})]{Betancourt2017}
Michael Betancourt.
\newblock How the shape of a weakly informative prior affects inferences,
  2017{\natexlab{a}}.
\newblock URL
  \url{https://mc-stan.org/users/documentation/case-studies/weakly_informative_shapes}.

\bibitem[Betancourt(2017{\natexlab{b}})]{Betancourt2018}
Michael Betancourt.
\newblock A {C}onceptual introduction to {H}amiltonian {M}onte {C}arlo,
  2017{\natexlab{b}}.
\newblock Preprint at \url{https://arxiv.org/abs/1701.02434v2}.

\bibitem[Buonaccorsi(2010)]{Buonaccorsi2010}
John~P Buonaccorsi.
\newblock \emph{Measurement error: models, methods, and applications}.
\newblock Chapman and Hall/CRC, Boca Raton, 2010.

\bibitem[Carpenter et~al.(2017)Carpenter, Gelman, Hoffman, Lee, Goodrich,
  Betancourt, Brubaker, Guo, Li, and Riddell]{Carpenter2017}
Bob Carpenter, Andrew Gelman, Matthew~D. Hoffman, Daniel Lee, Ben Goodrich,
  Michael Betancourt, Marcus Brubaker, Jiqiang Guo, Peter Li, and Allen
  Riddell.
\newblock Stan: A probabilistic programming language.
\newblock \emph{Journal of statistical software}, 76\penalty0 (1), 2017.
\newblock \doi{10.18637/jss.v076.i01}.

\bibitem[Carroll et~al.(2006)Carroll, Ruppert, Stefanski, and
  Crainiceanu]{Carroll2006}
Raymond~J. Carroll, David Ruppert, Leonard~A. Stefanski, and Ciprian~M.
  Crainiceanu.
\newblock \emph{{Measurement Error in Nonlinear Models}}.
\newblock Chapman and Hall/CRC, London, 2006.

\bibitem[Crainiceanu and Goldsmith(2010)]{Crainiceanu2010}
Ciprian~M. Crainiceanu and A.~Jeffrey Goldsmith.
\newblock {Bayesian Functional Data Analysis Using WinBUGS}.
\newblock \emph{Journal Of Statistical Software}, 32\penalty0 (11):\penalty0
  195, 2010.
\newblock \doi{10.1103/PhysRevLett.106.107404}.

\bibitem[Croon and van Veldhoven(2007)]{Croon2007}
Marcel~A. Croon and Marc~J.P.M. van Veldhoven.
\newblock {Predicting group-level outcome variables from variables measured at
  the individual level: A latent variable multilevel model}.
\newblock \emph{Psychological Methods}, 12\penalty0 (1):\penalty0 45--57, 2007.
\newblock \doi{10.1037/1082-989X.12.1.45}.

\bibitem[Daouda et~al.(2021)Daouda, Hocine, and Temime]{Daouda2021}
Oumou~Salama Daouda, Mounia~N. Hocine, and Laura Temime.
\newblock {Determinants of healthcare worker turnover in intensive care units:
  A micro-macro multilevel analysis}.
\newblock \emph{Plos One}, 16\penalty0 (5):\penalty0 1--13, 2021.
\newblock \doi{10.1371/journal.pone.0251779}.

\bibitem[Eilers and Marx(1996)]{Eilers1996}
Paul H.~C. Eilers and Brian~D. Marx.
\newblock {Flexible Smoothing with B-splines and Penalties}.
\newblock \emph{Statistical Science}, 11\penalty0 (2):\penalty0 89--121, 1996.
\newblock \doi{10.1214/ss/1038425655}.

\bibitem[Foster-Johnson and Kromrey(2018)]{FosterJohnson2018}
Lynn Foster-Johnson and Jeffrey~D. Kromrey.
\newblock {Predicting group-level outcome variables: An empirical comparison of
  analysis strategies}.
\newblock \emph{Behavior Research Methods}, 50\penalty0 (6):\penalty0
  2461--2479, 2018.
\newblock \doi{10.3758/S13428-018-1025-8}.

\bibitem[Gelman(2006)]{Gelman2006}
Andrew Gelman.
\newblock {Prior distributions for variance parameters in hierarchical models}.
\newblock \emph{Bayesian Analysis}, 1\penalty0 (3):\penalty0 515--533, 2006.
\newblock \doi{10.1214/06-BA117A}.

\bibitem[Gelman and Rubin(1992)]{Gelman1992}
Andrew Gelman and Donald~B. Rubin.
\newblock {Inference from Iterative Simulation Using Multiple Sequences}.
\newblock \emph{Statistical Science}, 7\penalty0 (4):\penalty0 457 -- 472,
  1992.
\newblock \doi{10.1214/ss/1177011136}.

\bibitem[Gelman et~al.(2013)Gelman, Carlin, Stern, Dunson, Vehtari, and
  Rubin]{Gelman2013}
Andrew Gelman, John~B. Carlin, Hal~S. Stern, David~B. Dunson, Aki Vehtari, and
  Donald~B. Rubin.
\newblock \emph{Bayesian Data Analysis}.
\newblock Chapman and Hall/CRC, Boca Raton, third edition, 2013.

\bibitem[Goldstein(2010)]{Goldstein2010}
Harvey Goldstein.
\newblock \emph{{Multilevel statistical models}}.
\newblock John Wiley \& Sons, Chichester, fourth edition, 2010.
\newblock ISBN 9780470973400.

\bibitem[Hansen(2014)]{Hansen2014}
Bruce~E. Hansen.
\newblock {5Nonparametric Sieve Regression: Least Squares, Averaging Least
  Squares, and Cross-Validation}.
\newblock In \emph{{The Oxford Handbook of Applied Nonparametric and
  Semiparametric Econometrics and Statistics}}. Oxford University Press, 02
  2014.
\newblock ISBN 9780199857944.
\newblock \doi{10.1093/oxfordhb/9780199857944.013.008}.

\bibitem[Hoffman and Gelman(2014)]{Hoffman2014}
Matthew~D. Hoffman and Andrew Gelman.
\newblock {The No-U-Turn Sampler: Adaptively Setting Path Lengths in
  Hamiltonian Monte Carlo}.
\newblock \emph{Journal of Machine Learning Research}, 15\penalty0
  (1):\penalty0 1593--1623, 2014.

\bibitem[Hsiao(1989)]{Hsiao1989}
Cheng Hsiao.
\newblock {Consistent estimation for some nonlinear errors-in-variables
  models}.
\newblock \emph{Journal of Econometrics}, 41\penalty0 (1):\penalty0 159--185,
  1989.
\newblock \doi{10.1016/0304-4076(89)90047-X}.

\bibitem[Hu and Schennach(2008)]{Hu2008}
Yingyao Hu and Susanne~M. Schennach.
\newblock {Instrumental Variable Treatment of Nonclassical Measurement Error
  Models}.
\newblock \emph{Econometrica}, 76\penalty0 (1):\penalty0 195--216, 2008.
\newblock \doi{10.1111/j.0012-9682.2008.00823.x}.

\bibitem[Lambert and Eilers(2006)]{Lambert2006}
Philippe Lambert and Paul H.~C. Eilers.
\newblock {Bayesian multi-dimensional density estimation with P-splines}.
\newblock In John Hinde, Jochen Einbeck, and John Newell, editors,
  \emph{Proceedings of the 21st International Workshop on Statistical
  Modelling}, pages 313--320, 2006.

\bibitem[Lang and Brezger(2004)]{Lang2004}
Stefan Lang and Andreas Brezger.
\newblock {Bayesian P-Splines}.
\newblock \emph{Journal of Computational and Graphical Statistics}, 13\penalty0
  (1):\penalty0 183--212, 2004.
\newblock \doi{10.1198/1061860043010}.

\bibitem[Li(2002)]{Li2002}
Tong Li.
\newblock {Robust and consistent estimation of nonlinear errors-in-variables
  models}.
\newblock \emph{Journal of Econometrics}, 110\penalty0 (1):\penalty0 1--26,
  2002.
\newblock \doi{0.1016/S0304-4076(02)00120-3}.

\bibitem[Li and Cao(2022)]{Li2022}
Zheyuan Li and Jiguo Cao.
\newblock {General P-Splines for Non-Uniform B-Splines}, 2022.
\newblock Preprint at \url{https://arxiv.org/abs/2201.06808v2}.

\bibitem[Lu et~al.(2017)Lu, Page-Gould, and Xu]{micromacro}
Jackson~G Lu, Elizabeth Page-Gould, and Nancy~R Xu.
\newblock \emph{MicroMacroMultilevel: Micro-Macro Multilevel Modeling}, 2017.
\newblock R package version 0.4.0.

\bibitem[McDonald(2022)]{mcdonald2022}
Shaun McDonald.
\newblock \emph{Novel approaches to uncertainty quantification in nonparametric
  settings}.
\newblock PhD thesis, 2022.

\bibitem[O'Sullivan(1988)]{OSullivan1988}
Finbarr O'Sullivan.
\newblock {Fast Computation of Fully Automated Log-Density and Log-Hazard
  Estimators}.
\newblock \emph{SIAM Journal on Scientific and Statistical Computing},
  9\penalty0 (2):\penalty0 363--379, 1988.
\newblock \doi{10.1137/0909024}.

\bibitem[Park and Casella(2008)]{Park2008}
Trevor Park and George Casella.
\newblock {The Bayesian Lasso}.
\newblock \emph{Journal of the American Statistical Association}, 103\penalty0
  (482):\penalty0 681--686, 2008.
\newblock \doi{10.1198/016214508000000337}.

\bibitem[{R Core Team}(2020)]{R}
{R Core Team}.
\newblock \emph{R: A Language and Environment for Statistical Computing}.
\newblock R Foundation for Statistical Computing, Vienna, Austria, 2020.

\bibitem[Ramsay and Siverman(2005)]{Ramsay2005}
James~O. Ramsay and Bernard~W. Siverman.
\newblock \emph{{Functional Data Analysis}}.
\newblock Springer Series in Statistics. Springer, New York, 2005.

\bibitem[Richardson and Gilks(1993)]{Richardson1993}
Sylvia Richardson and Walter~R. Gilks.
\newblock {Conditional Independence Models for Epidemiological Studies with
  Covariate Measurement Error}.
\newblock \emph{Statistics in Medicine}, 12\penalty0 (18):\penalty0 1703--1722,
  1993.
\newblock \doi{10.1002/sim.4780121806}.

\bibitem[Rousseau(2016)]{Rousseau2016}
Judith Rousseau.
\newblock {On the Frequentist Properties of Bayesian Nonparametric Methods}.
\newblock \emph{The Annual Review of Statistics and Its Applications},
  3:\penalty0 211--231, 2016.
\newblock \doi{10.1146/annurev-statistics-041715-033523}.

\bibitem[Sarkar et~al.(2014)Sarkar, Mallick, and Carroll]{Sarkar2014}
Abhra Sarkar, Bani~K. Mallick, and Raymond~J. Carroll.
\newblock {Bayesian Semiparametric Regression in the Presence of Conditionally
  Heteroscedastic Measurement and Regression Errors}.
\newblock \emph{Biometrics}, 70\penalty0 (4):\penalty0 823--834, 2014.
\newblock \doi{10.1111/biom.12197}.

\bibitem[Schennach(2016)]{Schennach2016}
Susanne~M. Schennach.
\newblock {Recent Advances in the Measurement Error Literature}.
\newblock \emph{Annual Review of Economics}, 8:\penalty0 341--377, 2016.
\newblock \doi{10.1146/annurev-economics-080315-015058}.

\bibitem[Serra and Krivobokova(2017)]{Serra2017}
Paulo Serra and Tatyana Krivobokova.
\newblock {Adaptive Empirical Bayesian Smoothing Splines}.
\newblock \emph{Bayesian Analysis}, 12\penalty0 (1):\penalty0 219 -- 238, 2017.
\newblock \doi{10.1214/16-BA997}.

\bibitem[Silverman(1982)]{Silverman1982}
B.W. Silverman.
\newblock {On the Estimation of a Probability Density Function by the Maximum
  Penalized Likelihood Method}.
\newblock \emph{The Annals of Statistics}, 10\penalty0 (3):\penalty0 795--810,
  1982.
\newblock \doi{10.1214/aos/1176345872}.

\bibitem[Snijders and Bosker(2011)]{Snijders2011}
Tom A.~B. Snijders and Roel~J. Bosker.
\newblock \emph{Multilevel analysis: An introduction to basic and advanced
  multilevel modeling}.
\newblock SAGE Publications Ltd., Los Angeles, 2011.

\bibitem[{Stan Development Team}(2018)]{stanref}
{Stan Development Team}.
\newblock Stan modeling language users guide and reference manual, version
  2.30.0, 2018.

\bibitem[{Stan Development Team}(2021)]{rstan}
{Stan Development Team}.
\newblock \emph{{RStan}: the {R} interface to {Stan}}, 2021.
\newblock R package version 2.21.3.

\bibitem[van~de Wiel et~al.(2019)van~de Wiel, {Te Beest}, and
  M{\"{u}}nch]{VandeWiel2019}
Mark~A. van~de Wiel, Dennis~E. {Te Beest}, and Magnus~M. M{\"{u}}nch.
\newblock {Learning from a lot: Empirical Bayes for high-dimensional
  model-based prediction}.
\newblock \emph{Scandinavian Journal of Statistics}, 46\penalty0 (1):\penalty0
  2--25, 2019.
\newblock \doi{10.1111/sjos.12335}.

\bibitem[Varanasi and Aazhang(1989)]{Varanasi1989}
Mahesh~K. Varanasi and Behnaam Aazhang.
\newblock Parametric generalized gaussian density estimation.
\newblock \emph{The Journal of the Acoustical Society of America}, 86\penalty0
  (4):\penalty0 1404--1415, 1989.
\newblock \doi{10.1121/1.398700}.

\bibitem[Vehtari et~al.(2021)Vehtari, Gelman, Simpson, Carpenter, and
  B{\"u}rkner]{Vehtari2021}
Aki Vehtari, Andrew Gelman, Daniel Simpson, Bob Carpenter, and Paul-Christian
  B{\"u}rkner.
\newblock {Rank-normalization, folding, and localization: an improved $\hat{R}$
  for assessing convergence of MCMC (with discussion)}.
\newblock \emph{Bayesian analysis}, 16\penalty0 (2):\penalty0 667--718, 2021.
\newblock \doi{10.1214/20-BA1221}.

\end{thebibliography}

\end{document}